\documentclass[UTF8]{article}
\usepackage{amsmath}
\usepackage{graphicx}
\usepackage{float}
\usepackage{graphicx,subfigure}

\usepackage[left=1cm,right=1cm,top=3cm,bottom=3cm]{geometry}

\usepackage{a4wide}
\usepackage{amsmath}
\usepackage[
      colorlinks=true,
      linkcolor=blue,
      urlcolor=blue,
      filecolor=black,
      citecolor=blue,
            ]{hyperref}
\usepackage{color}
\usepackage{amsfonts}
\usepackage{amssymb}
\usepackage{epstopdf}

\begin{document}

\title{\Large \textbf{No Inner-Horizon Theorem for Black Holes with Charged Scalar Hairs}}
\author{\large
Rong-Gen Cai$^{a,b,c}$\footnote{cairg@itp.ac.cn}~,
~~Li Li$^{\,a,b,c}$\footnote{liliphy@itp.ac.cn (corresponding author)}~,
~~Run-Qiu Yang$^{d}$\footnote{aqiu@tju.edu.cn  (corresponding author)}
\\
\\
\small $^a$CAS Key Laboratory of Theoretical Physics, Institute of Theoretical Physics, \\
\small Chinese Academy of Sciences, Beijing 100190, China\\
\small $^b$School of Physical Sciences, University of Chinese Academy of Sciences, \\
\small No.19A Yuquan Road, Beijing 100049, China\\
\small $^c$School of Fundamental Physics and Mathematical Sciences, Hangzhou Institute for Advanced Study, \\
\small UCAS, Hangzhou 310024, China\\
\small $^d$Center for Joint Quantum Studies and Department of Physics, School of Science, Tianjin University, \\
\small Yaguan Road 135, Jinnan District, 300350 Tianjin, China
}
\date{}

\maketitle

\begin{abstract}
We establish a no inner-horizon theorem for  black holes with charged scalar hairs. Considering a general gravitational theory with a charged scalar field, we prove that there exists no inner Cauchy horizon for both spherical and planar black holes with non-trivial scalar hair. The hairy black holes approach  to a spacelike singularity at late interior time. This result is independent of the form of scalar potentials as well as the asymptotic boundary of spacetimes.  We prove that the geometry near the singularity takes a universal Kasner form when the kinetic term of the scalar hair dominates, while novel behaviors different  from the Kasner form are uncovered when the scalar potential become important to the background. For the hyperbolic horizon case, we show that hairy black hole can only has at most one inner horizon, and a concrete example with an inner horizon is presented. All these features are also valid for the Einstein gravity coupled with neutral scalars.
\end{abstract}

\newpage
\tableofcontents

\section{Introduction}

As one of the most fantastic objects among all gravitational compact objects, black holes play a central role in understanding the nature of gravity. The development of black hole physics has uncovered a deep and intrinsic relationship between gravitation, thermodynamics and quantum theory and has provided most of our present physical understandings of the quantum phenomena in strong gravity regime. In recent years there has been dramatic progress in understanding black hole physics both from theoretical and experimental aspects. In particular, thanks to the innovation and progress of observation techniques, one is able to directly detect the gravitational waves from a binary black hole coalescence~\cite{GWs:2016} and to take a photo of the shadow of a black hole~\cite{shadow:2019I,shadow:2019IV}, opening a new window in the study of gravity, astrophysics and cosmology.

While the exterior physics of black hole has been extensively investigated in the literature, in particular, the establishment of black hole thermodynamics and uniqueness theorems (see \emph{e.g.} Refs.~\cite{Wald:1999vt} and~\cite{Hollands:2012xy}, for reviews),  the interior structure of black holes behind  event horizon of black hole has not been well understood. Nevertheless, exploring the internal structure of black holes and spacetime singularities  inside are intriguing and fundamental topics in general relativity and can provide a better understanding of black hole physics,  gravitation and quantum physics. For example, the existence of the inner (Cauchy) horizon violates the predicability in general relativity and motivates the strong cosmic censorship (SCC) conjecture (see \emph{e.g.} Refs.~\cite{Ringstrom:2015jza,Isenberg:2015rqa}). Recent progress suggested that the black hole information paradox could be solved by including ``island" that lies in the interior of black hole~\cite{Almheiri:2019yqk,Almheiri:2019psy}. The possible observation of information from the interior of black holes in the future is of great significance to understanding the nature of our universe.

The most studied theory in general relativity involves a Maxwell field, for which Schwarzschild, Reissner-Nordstr\"{o}m (RN)  and Kerr-Newmann black holes are three well-known solutions. The neutral Schwarzschild black hole has an event horizon and a spacelike singularity inside, while the later two have one additional Cauchy horizon that appears to violate SCC and a timelike singularity at the center, due to the presence of non-trivial electric charge and angular momentum. On the other hand, scalar fields should be one of the simplest types of ``matter" considered in the literature and plays important role in particle physics, cosmology and gravitational physics.  Due to their simplicity, it is quite natural to consider scalar fields when testing some no-go ideas as a first step. Recently, it has been argued that SCC can be violated by turning on linear scalar field perturbations of RN black holes in de Sitter (dS) space~\cite{Cardoso:2017soq}.

One anticipates that the black hole interior would be dramatically affected in presence of scalar hair. Indeed, it has been recently shown that~\cite{Hartnoll:2020rwq} there is no inner Cauchy horizon for some kind of charged black holes with a neutral scalar. However, the result relies on a strong requirement for the scalar potential (the scalar mass-square should be negative) and breaks down for charged scalar case (the absence of inner horizon for planar black holes with charged scalar was discussed in Ref.~\cite{Hartnoll:2020fhc} more recently, see \textbf{Note added} also). In this work we will establish a stronger theorem for the inner structure of black holes for both the Einstein-scalar and the Einstein-Maxwell-charged scalar theories. Our results are quite generic and are independent of the form of scalar potentials as well as the asymptotic geometry of spacetime. We will also discuss the asymptotical solutions near the singularity and uncover some new features. In addition to the Kasner form of solutions for which the kinetic terms dominate the dynamics, we will show numerical evidence for the existence of novel oscillating behaviors all the way down to the singularity when the potential terms become important to the geometry.

\section{The Model}
We consider a $(d+2)$-dimensional theory with gravity coupled with a Maxwell field $A_\mu$ and a charged scalar field $\Psi$:
\begin{eqnarray}
\label{action}S&=&\frac{1}{2\kappa_N^2}\int d^{d+2}x \sqrt{-g} \left[\mathcal{R}+\mathcal{L}_M\right]\,, \\
\mathcal{L}_M&=&-\frac{Z(|\Psi|^2)}{4}F_{\mu\nu}F^{\mu\nu}-(D_\mu\Psi)^* D^{\mu}\Psi-V(|\Psi|^2)\,,\nonumber
\end{eqnarray}
where  $F_{\mu\nu}=\nabla_\mu A_\nu-\nabla_\nu A_\mu$ and  $D_\mu=\nabla_\mu -i q A_\mu$ with $q$ the charge of the scalar field. $Z$ and $V$ are arbitrary smooth functions of $|\Psi|^2$. One only demands $Z$ to be positive to ensure positivity of the kinetic term for $A_\mu$, and takes $Z(0)=1$ without loss of generality. The Einstein-scalar theory is obtained by turning off $A_\mu$ and most of our discussion below will apply to black holes in the Einstein-scalar case\,\footnote{Although there are some no-scalar-hair theorems for the existence of black hole solutions  sourced by a non-trivial scalar field in Einstein-scalar gravity~\cite{Herdeiro:2015waa,Bhattacharya:2007ap}, static black holes with non-trivial neutral scalar hair do exist, see \emph{e.g.}~\cite{Dennhardt:1996cz,Zloshchastiev:2004ny,Bronnikov:2005gm,Cadoni:2015gfa,Feng:2013tza,Ren:2019lgw,Li:2020spf}.}. Note that the spacetime can be asymptotically flat, anti-de Sitter (AdS), dS or other geometries, depending on the choice of the scalar potential $V$ as well as the coupling $Z$.

The equations of motion from the action are
\begin{equation}\label{smEOMs}
\begin{split}
&D_\mu D^\mu \Psi -\left(\dot{V}(|\Psi|^2)+\frac{\dot{Z}(|\Psi|^2)}{4}F_{\mu\nu}F^{\mu\nu}\right)\Psi=0 \,,  \\
&\nabla^\mu[Z(|\Psi|^2)  F_{\mu\nu} ]= i q ( \Psi^* D_\nu \Psi - \Psi D_\nu \Psi^*   ) \,,  \\
& \mathcal{R}_{\mu\nu} -\frac{1}{2}\mathcal{R}g_{\mu\nu}=\frac{1}{2}\mathcal{L}_M g_{\mu\nu}+ \frac{Z(|\Psi|^2)}{2}F_{\mu\sigma}{F_\nu}^\sigma+ \frac{1}{2}\left[ D_\mu \Psi (D_\nu \Psi)^* + D_\nu \Psi (D_\mu \Psi)^* \right]\,.
\end{split}
\end{equation}
We are interested in static charged black holes that are homogeneous and isotropic, so that the ansatz for metric and matter fields can be written as
\begin{equation}\label{ansatz}
\begin{split}
ds^2=\frac{1}{z^2}\left[-f(z)e^{-\chi(z)}dt^2+\frac{dz^2}{f(z)}+d\Sigma^2_{d,k}\right]\,,\\
\Psi=\psi(z)\,\quad A=A_t(z) dt\,,
\end{split}
\end{equation}
where the $d$-dimensional line element $d\Sigma_{d,k}^2$ is
\begin{equation}\label{Sigmak}
  d \Sigma_{d,k}^2=\left\{
  \begin{split}
  &d\theta^2+\sin^2\theta d\Omega_{d-1}^2,~~~~k=1\,,\\
  &\sum_{i=1}^{d} d x^2_i,~~~~~~~~~~~~~~~~~k=0\,,\\
  &d\theta^2+\sinh^2\theta d\Omega_{d-1}^2,~~k=-1\,,
  \end{split}
  \right.
\end{equation}
with $d\Omega_{d-1}^2$ the line element of $(d-1)$-dimensional unit sphere. We assume that the black hole boundary is at $z=0$ and the singularity at $z\rightarrow\infty$, but the precise location of the boundary and singularity is not important in our discussion. For black hole spacetimes obeying the dominant energy condition only the spherical horizon  case is allowed~\cite{Galloway:2005mf}. The well known case that breaks the dominant energy condition is to introduce a negative cosmological constant, for which the topology of a black hole horizon can be flat or hyperbolic~\cite{Lemos:1994xp,Cai:1996eg,Mann:1996gj,Brill:1997mf,Vanzo:1997gw}. In particular, the planar case has been widely investigated in the application of holographic duality to strongly coupled systems~\cite{Jan:book,Ammon:book,Cai:2015cya,Hartnoll:2016apf,Baggioli:2021xuv}.

With ansatz~\eqref{ansatz}, we obtain the following equations of motion:
\begin{eqnarray}
z^{d+2}e^{\chi/2}(e^{-\chi/2}z^{-d}f\psi')'&=&\left[\dot{V}_{\text{eff}}(\psi^2)-\frac{q^2 z^2 A_t^2 e^{\chi}}{f}\right]\psi \label{smeompsi}\,,\\
z^{d}[Z(\psi^2) e^{\chi/2}z^{2-d}A_t']'&=&\frac{2 q^2 \psi^2 e^{\chi/2}}{f}A_t\label{smeomAt}\,, \\
\frac{d}{2}\chi' &=&z\psi'^2+\frac{z e^{\chi} q^2 \psi^2A_t^2}{f^2}\label{smeomchi}\,,\\
\frac{d}{2}\frac{f'}{f}-\frac{z}{2}\psi'^2-\frac{d(d+1)}{2 z} &=&\frac{V_{\text{eff}}(\psi^2)}{2 z f}-\frac{k d(d-1) z}{2f}
+\frac{z e^{\chi}q^2A_t^2}{2f^2}\psi^2+\frac{Z(\psi^2)z^3 e^{\chi}A_t'^2}{2 f}\label{smeomf}\,,
\end{eqnarray}
where a prime denotes the derivative with respect to $z$ and $V_{\text{eff}}(x)=V(x)-\frac{1}{2}Z(x)z^4 e^{\chi}A_t'^2$ with $\dot{V}_{\text{eff}}(x)=\mathrm{d} V_{\text{eff}}(x)/\mathrm{d}x$. Without loss of generality, we can choose $\psi(z)$ to be real because of the symmetry in the problem. In general, the above coupled equations do not have analytical solutions, so one has to solve the system numerically. Eqs.~\eqref{smeompsi}-\eqref{smeomf} have been solved numerically outside black holes ($z<z_H$) in the literature. To continue behind the event horizon it is simple to switch to ingoing coordinate, for which the equations of motion do not change.

\begin{figure}[h!]
\begin{center}
\includegraphics[width=0.7\textwidth]{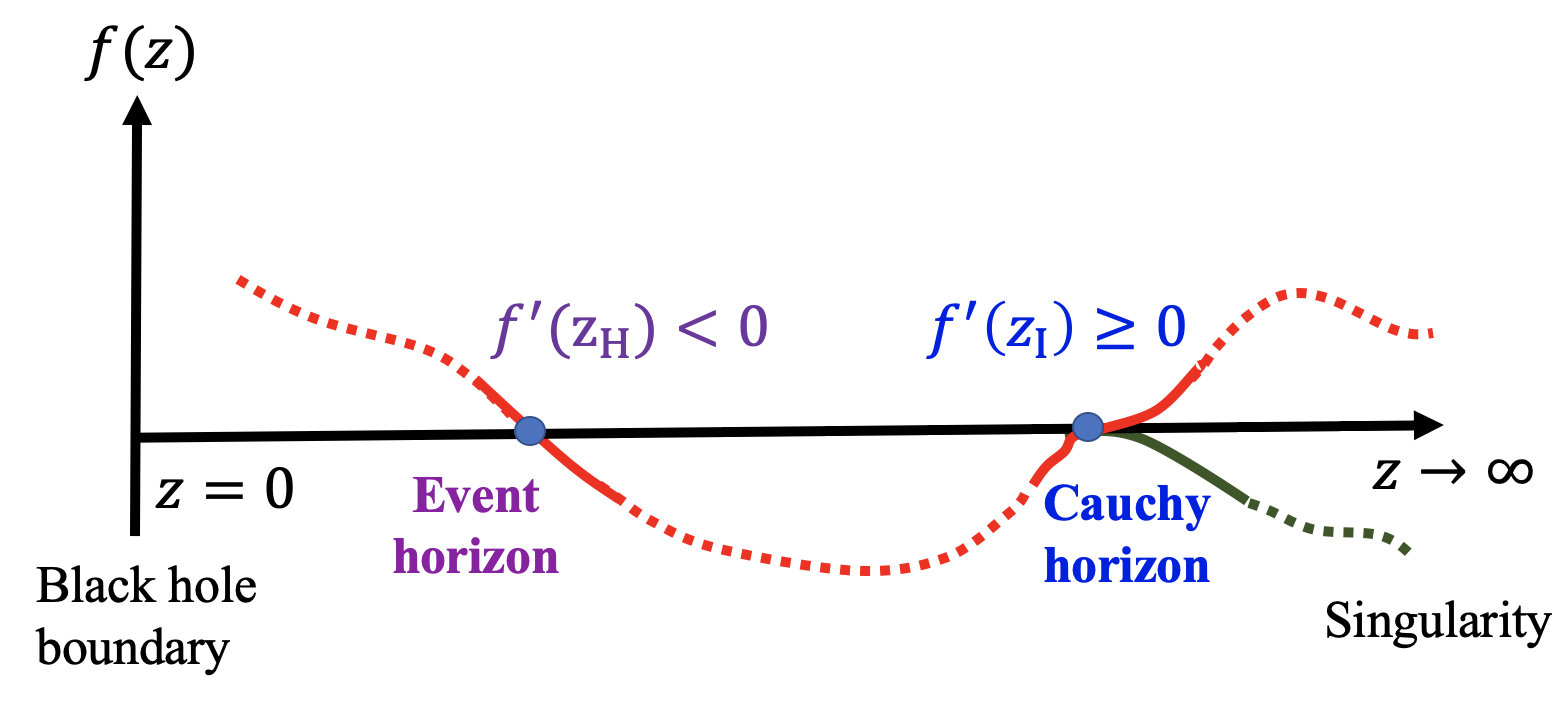}
\caption{The schematic structure of a charged black hole with two horizons. The blackening function $f(z)$ vanishes both at the event horizon $z=z_H$ and the inner horizon $z=z_I$.}
\label{fig:BH}
\end{center}
\end{figure}

\section{Proof of No Inner-Horizon}
Suppose that there were two horizons, including the event horizon at $z_H$ and an inner horizon at $z_I$, for which $f(z_H)=f(z_I)=0$ with $z_H<z_I$. In the present study we consider black holes with finite temperature. The structure of the black hole is shown schematically in Fig.~\ref{fig:BH}. The blackening function $f(z)$ turns from positive to negative towards the interior near $z_H$, while it from negative to positive near the inner horizon $z_I$ (red curve) or $z_I$ is a local maximum (green curve).  Therefore, one has
\begin{equation}\label{constf}
f'(z_H)<0,\quad f'(z_I)\geq0\,.
\end{equation}
Furthermore, to have a regular horizon, the metric and matter fields should be sufficiently smooth near the horizon. For the black hole solution with non-trivial charged scalar $\psi$, the equations of motion  imply  the condition:
\begin{equation}\label{constAt}
A_t(z_H)=A_t(z_I)=0\,,
\end{equation}
with both $\psi$ and $\chi$ finite at two horizons.

Before going to prove the no inner-horizon theorem, we briefly show why the  discussion of Ref.~\cite{Hartnoll:2020rwq} does not work for the charged scalar case. Following Ref.~\cite{Hartnoll:2020rwq}, we obtain from the scalar equation that
\begin{equation}\label{scalarBL}
\begin{split}
0=\int_{z_H}^{z_I}\left(\frac{f e^{\chi/2}\psi\psi'}{z^d}\right)'=\int_{z_H}^{z_I} \frac{e^{-\chi/2}}{z^{d+2}}\left[\psi^2 \dot{V}_{\text{eff}} +z^2 f\psi'^2-\frac{q^2 z^2 e^{\chi}A_t^2\psi^2}{f}\right],
\end{split}
\end{equation}
where a prime denotes the derivative with respect to $z$ and the effective potential $V_{\text{eff}}(x)=V(x)-\frac{1}{2}Z(x)z^4 e^{\chi}A_t'^2$ with $\dot{V}_{\text{eff}}(x)=\mathrm{d} V_{\text{eff}}(x)/\mathrm{d}x$. For the neutral case with $q=0$, the integrand in the second line  is non-positive over the range $(z_H, z_I)$ provided $\dot{V}_{\text{eff}}<0$\,\footnote{The authors of Ref.~\cite{Hartnoll:2020rwq} considered a neutral scalar with $V=m^2 \psi^2$ and $Z=1$, which yields $\dot{V}_{\text{eff}}<0$ for $m^2 <0$.}. Therefore, the only way for two horizons is for $\psi=0$. However, for the charged scalar case with non-zero $q$, there is an additional contribution which is positive. So we cannot rule out the existence of inner horizon even when $\dot{V}_{\text{eff}}<0$. Nevertheless, we will show that there is no inner horizon for a charged black hole with the curvature of its horizon  be non-negative. The hyperbolic case is a bit complicated, but we are able to show that it has at most one inner horizon with non-vanishing surface gravity.

Our key observation for the background~\eqref{ansatz} is the existence of the conserved quantity
\begin{equation}\label{myQ}
\mathcal{Q}(z)=z^{2-d}e^{\chi/2}\left[z^{-2}(fe^{-\chi})'-ZA_t A_t'\right]+2k (d-1)\int^z y^{-d}e^{-\chi(y)/2}dy\,,
\end{equation}
for which $\mathcal{Q}'(z)=0$. For the planar case with $k=0$, $\mathcal{Q}$ was constructed in the literature (see \emph{e.g.} Refs.~\cite{Gubser:2009cg,Kiritsis:2015hoa}) and is due to a particular scaling symmetry that is only valid for the planar topology~\cite{Gubser:2009cg}. Intriguingly, we manage to find a radially conserved $\mathcal{Q}$ in Eq.~\eqref{myQ} for non-planar cases even the scaling symmetry breaks down.

Evaluating $\mathcal{Q}$ both at the event and inner horizons, we obtain
\begin{equation}
\mathcal{Q}(z_j)=\frac{f'(z_j)}{z_j^{d}}e^{-\chi(z_j)/2}+2k(d-1)\int^{z_j} y^{-d}e^{-\chi(y)/2}dy\,,
\end{equation}
where the subscript $j=(H, I)$ and we have used Eq.~\eqref{constAt}. Since $\mathcal{Q}(z_H)=\mathcal{Q}(z_I)$, we have
\begin{equation}\label{minusQ}
\frac{f'(z_H)}{z_H^{d}}e^{-\chi(z_H)/2}-\frac{f'(z_I)}{z_I^{d}}e^{-\chi(z_I)/2}=2k(d-1)\int_{z_H}^{z_I}y^{-d}e^{-\chi(y)/2}dy\,.
\end{equation}
It is obvious that the left hand side is negative because of Eq.~\eqref{constf}. For black holes with spherical ($k=1$) and planar ($k=0$) topologies, the right hand side of Eq.~\eqref{minusQ} is non-negative. Therefore, smooth inner Cauchy horizon is never  able to form for spherical and planar black holes with charged scalar hair.

\begin{figure}[h!]
\begin{center}
\includegraphics[width=0.5\textwidth]{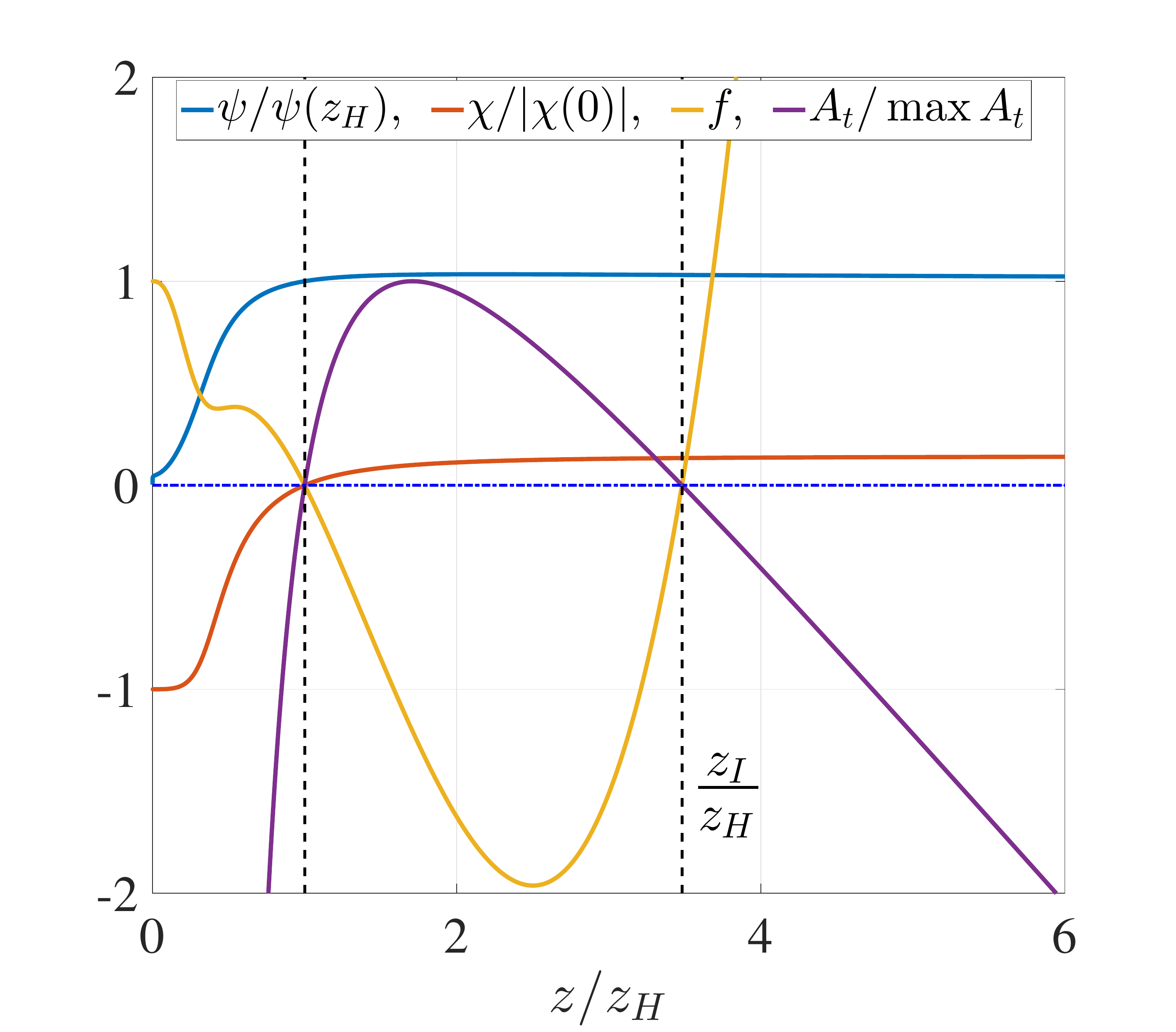}
\caption{Numerical solution for hyperbolic black hole in AdS. The hairy black hole has an event horizon at $z_H=1.193936$ and a Cauchy horizon at $z_I\approx4.15699837$. We have considered the four dimensional model with $V=-6-0.18388\psi^2, Z=1$ and $q=1.5$. }
\label{fig:hyperbolic}
\end{center}
\end{figure}
For the hyperbolic case with $k=-1$, since both sides of Eq.~\eqref{minusQ} share the same sign, it is possible to develop an inner horizon, provided Eqs.~\eqref{scalarBL} and~\eqref{minusQ} are satisfied. A concrete example for the hyperbolic black hole with an inner horizon is presented in Fig.~\ref{fig:hyperbolic} (see Appendix~\ref{HBexample} for numerical details).
Nevertheless, we can show that for the hyperbolic case, there only exists at most one inner horizon with nonzero surface gravity. The proof contains two steps.  We first prove the horizon $z_I$ must be a single root. Otherwise, we must have $f''(z_I)\leq0$. Computing  $\mathcal{Q}'(z_I)$, we find
\begin{equation}
\mathcal{Q}'(z_I)=\frac{f''(z_I)}{z_I^{d}}e^{-\chi(z_I)/2}-Zz_I^{2-d}e^{\chi(z_I)/2}A_t'(z_I)^2-2(d-1)z_I^{-d}e^{-\chi(z_I)/2}<0\,.
\end{equation}
This is contradictory to the fact that $\mathcal{Q}'(z)=0$.  Thus, $z_I$ should be a single root with $f'(z_I)>0$.
Secondly, suppose there is a second inner horizon appearing at $z=z_{II}>z_I$. It is then obvious that $f'(z_{II})\leq0$ with $f(z_{II})=0$. Using a similar discussion, we obtain

\begin{equation}\label{km1Q}
\frac{f'(z_{I})}{z_{I}^{d}}e^{-\chi(z_I)/2}-\frac{f'(z_{II})}{z_{II}^{d}}e^{-\chi(z_{II})/2}
=-2(d-1)\int_{z_{I}}^{z_{II}}y^{-d}e^{-\chi(y)/2}dy\,,
\end{equation}
for $k=-1$. While the right hand side is negative, the left hand side is positive, since $f'(z_I)>0$ and $f'(z_{II})\leq0$. Thus, the second inner horizon cannot appear.

\section{Geometry Near Singularity}
After knowing the inner structure behind the event horizon, we are now interested in the dynamics near the singularity. Since there involves the dynamics in nonlinear regimes, the behavior should be in general sensitive to the details of the model~\cite{Hartnoll:2020rwq,Hartnoll:2020fhc}. Instead of dealing with specific models, we aim to provide some generic features of the geometry near the singularity. Now, we assume that the singularity appears at $z\to \infty$.

To characterize the charge degrees of freedom behind the surface generating  a nonzero electric flux in the deep interior, we introduce~\cite{Gouteraux:2012yr}
\begin{equation}
Q(z)=\frac{1}{2\kappa_N^2}\int_{\Sigma} Z\, {}^\star F=-\frac{\omega_{(d)}}{2\kappa_N^2} Z z^{2-d}e^{\chi/2}A_t'\,,
\end{equation}
where $\omega_{(d)}$ is the volume of the section with $t$ and $z$ fixed and we have used the ansatz of Eq.~\eqref{ansatz}. For the hyperbolic case it is possible to have an inner horizon (see Fig.~\ref{fig:BH}). We find that this timelike singularity of a charged hyperbolic black hole always carries charge, \emph{i.e.} $Q(z\to\infty)\neq 0$. More precisely, one can prove that behind the inner horizon $z_I$, $Q(z)^2$ is  monotonically increasing towards the singularity  (see Appendix~\ref{gauge} for several lemmas of the gauge sector).

For other cases we have shown that there exists no inner horizon and the spacetime ends at a spacelike singularity. The geometry near the singularity depends on the details of a model one considers. We shall specify $Z=1$ to simplify the discussion. For the model in which the kinetic term of scalar dominants the dynamics, we can neglect the potential $V(\psi^2)$ and then obtain\,\footnote{See Appendix~\ref{SLsingularity} for more details.}
\begin{equation}\label{asyfchiaphi}
\begin{split}
\psi=\sqrt{d}\alpha\ln z+\cdots, \;A_t'=E_{s} z^{d-2-\alpha^2}+\cdots\,,\\
e^{\chi}=\chi_{s}z^{2\alpha^2}+\cdots,\;  f=-f_{s}z^{1+d+\alpha^2}+\cdots\,,
\end{split}
\end{equation}
as $z\to \infty$, with $(\alpha, E_s, \chi_s, f_s)$ constants. One can check that $Q(z\to\infty)$ approaches to a constant, so the spacelike singularity in the present case carries a finite charge.
Changing the $z$ coordinate to the proper time $\tau$ via $\tau\sim z^{-(1+d+\alpha^2)/2}$, we then obtain
\begin{equation}\label{kasner}
ds^2=-d\tau^2+c_t \tau^{2 p_t} d t^2+c_{s} \tau^{2 p_{s}}d\Sigma^2_{d,k}, \quad  \psi=-p_{\psi} \ln \tau\,,
\end{equation}
where
\begin{equation}
p_t=\frac{1-d+\alpha^2}{1+d+\alpha^2}, \; p_s=\frac{2}{1+d+\alpha^2}, \; p_{\psi}=\frac{2\sqrt{d}\alpha}{1+d+\alpha^2}\,.
\end{equation}
One immediately finds that
\begin{equation}
p_t+d p_s=1,\quad p_{t}^2+d p_{s}^2+p_{\psi}^2=1\,,
\end{equation}
and therefore the geometry has the Kasner form~\cite{Belinski:1973zz,Kasner:1921zz}.\footnote{A significant amount of work has been done for discussing Kasner-like behavior in the vicinity of a spacelike singularity, see \emph{e.g.}~\cite{Damour:2002tc,Belinski:2017fas} and references therein.}

\begin{figure}[h!]
\begin{center}
\subfigure[]{\includegraphics[width=.45\textwidth]{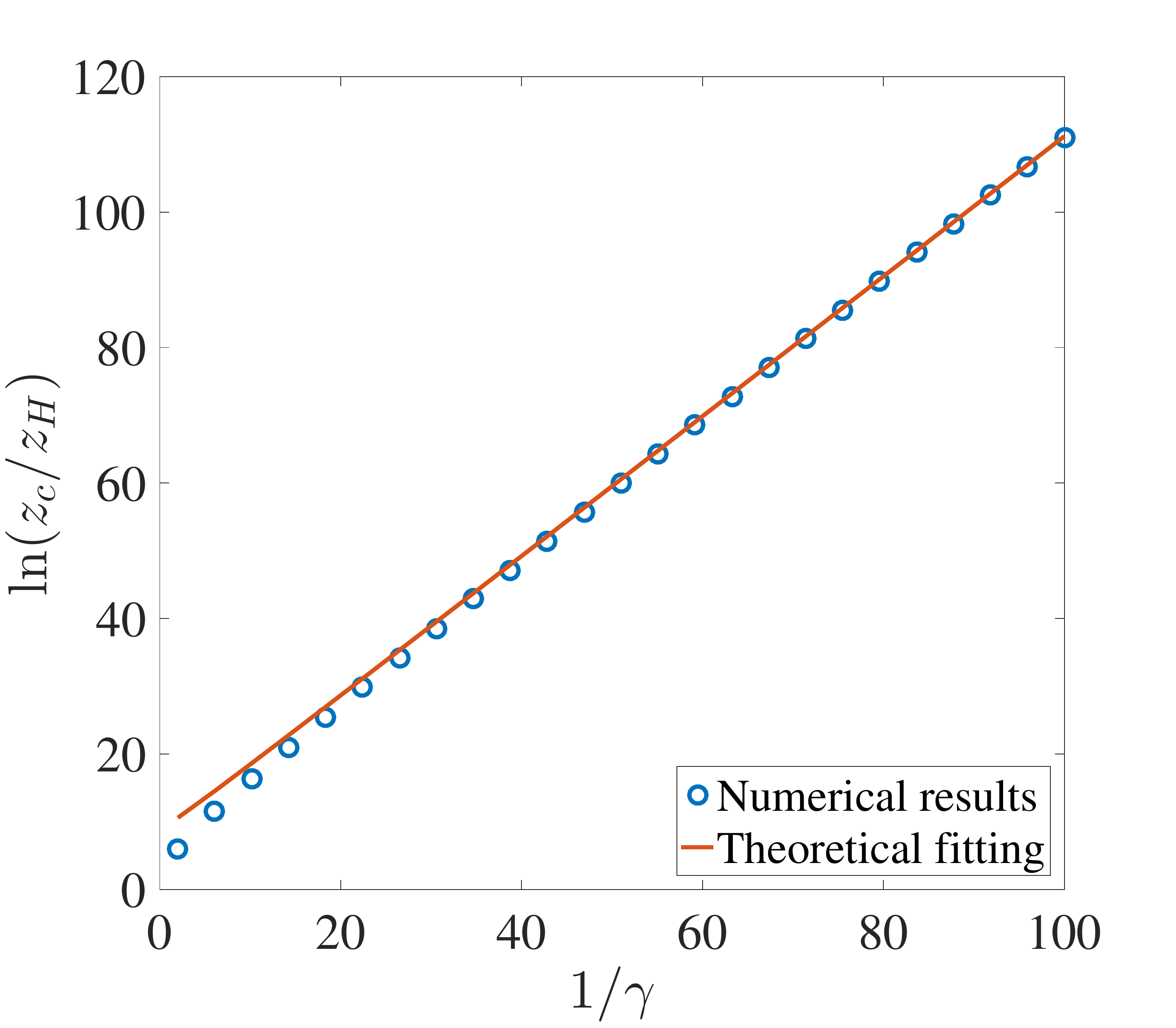}}
\subfigure[]{\includegraphics[width=.45\textwidth]{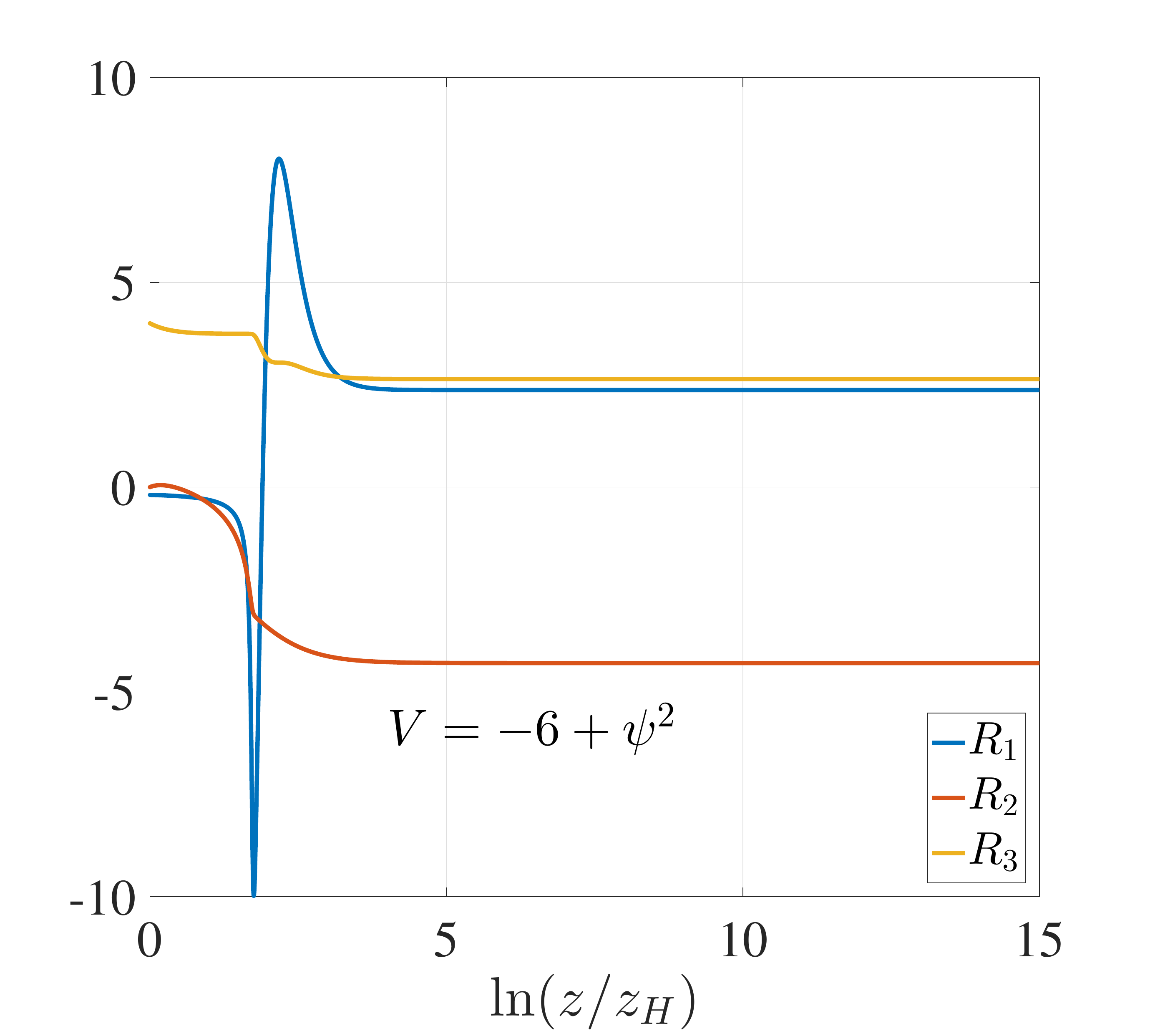}}
\subfigure[]{\includegraphics[width=.45\textwidth]{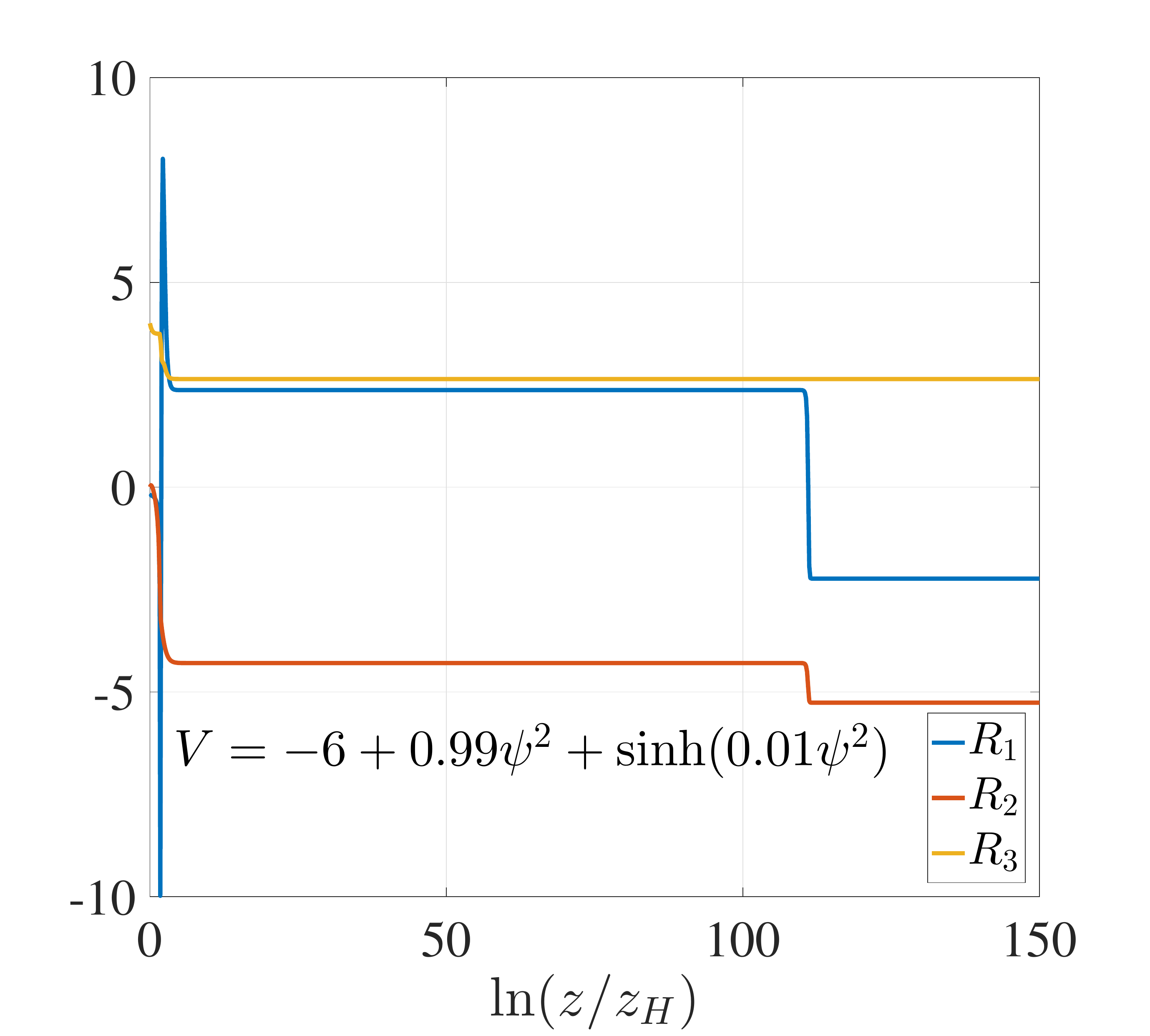}}
\subfigure[]{\includegraphics[width=.45\textwidth]{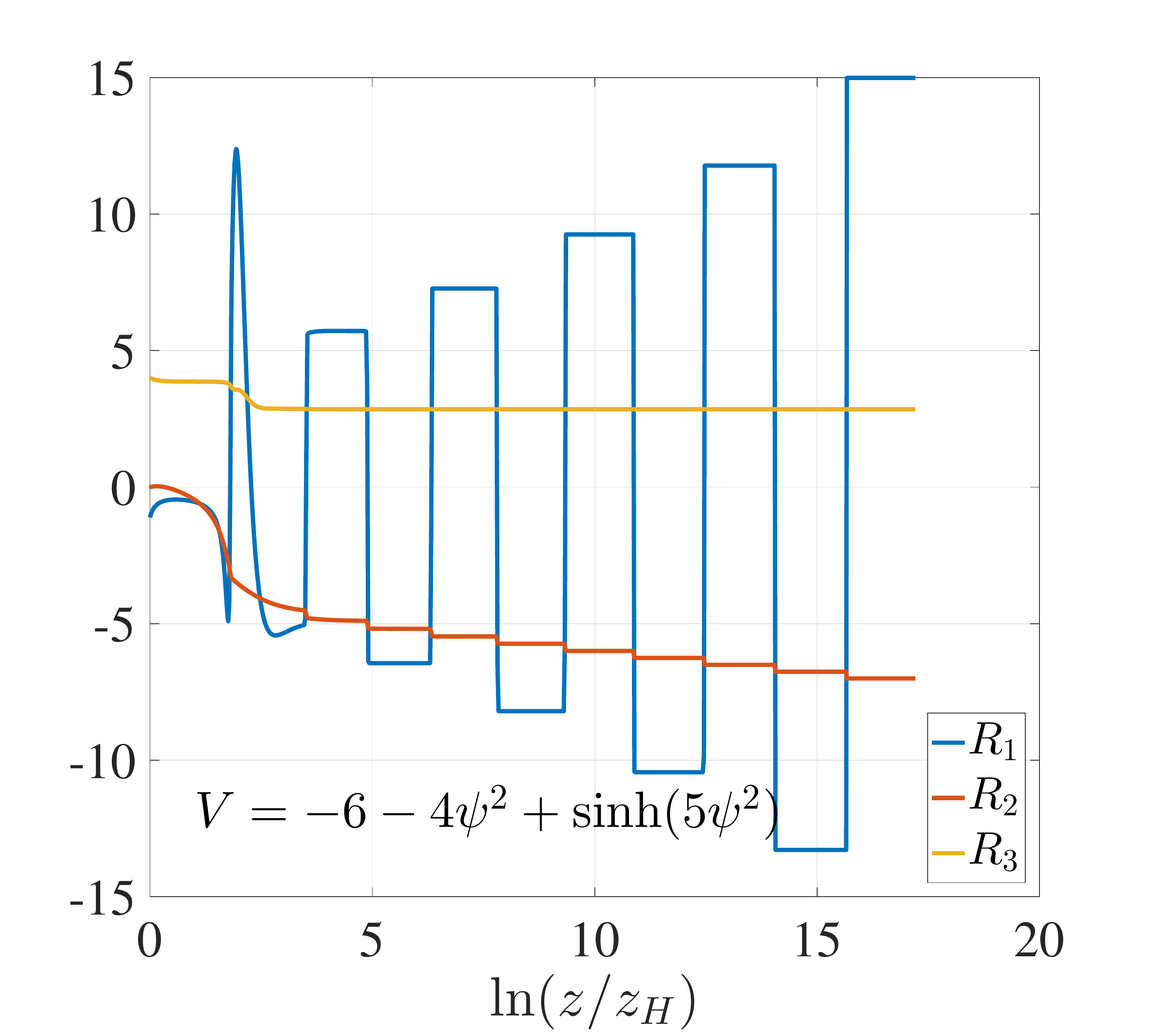}}
\caption{Numerical results for scalar potentials of Eq.~\eqref{exampvads}. Relationship between $z_c$ beyond which the Kasner behavior is modified and $\gamma$ matches with the theoretical prediction~\eqref{zcscaling} quite well, as shown in $(a)$. For polynomial potentials $(b)$, the numerical results satisfy the Kasner form. For non-polynomial potentials, the asymptotic solutions are different from the Kasner behavior. We observe strong oscillating behavior all the way down to the singularity, as shown in $(c)$ and $(d)$. We consider the planar black holes and specify $d=2, q=1$.}
\label{fig:singularity}
\end{center}
\end{figure}

As we have emphasized, for above discussion to be consistent, one should require the kinetic term of scalar to be dominant. In particular, one should at least has the following constraint:
\begin{equation}\label{constV1}
  \lim_{z\rightarrow\infty}\frac{|V(\psi^2)|}{z^{d+1+\alpha^2}}\ll 1\,,
\end{equation}
which allows the scalar potential $V$ to be arbitrary algebraic functions, including polynomial functions. For a potential that diverges exponentially or even worse, the condition~\eqref{constV1} is violated and the Kasner form~\eqref{kasner} would break down. For example, we take
\begin{equation}\label{exampvads}
  V=-6+(1-\gamma)\psi^2+\sinh(\gamma\psi^2)\,,
\end{equation}
for which the hairy black hole is asymptotically AdS. When $\gamma=0$, we expect to obtain~\eqref{asyfchiaphi}. In contrast, once turning on $\gamma>0$, the asymptotic solution would be different from the Kasner solution. The deviation from the Kasner form happens beyond a critical point
\begin{equation}\label{zcscaling}
z_c=c_0\, e^{\frac{d+1+\alpha^2}{\gamma d \alpha^2}}\,,
\end{equation}
with $c_0$ a coefficient (see Appendix~\ref{SLsingularity}). This scaling behavior is confirmed by our numerics in Fig.~\ref{fig:singularity}(a).

We also show some representative cases inside the event horizon for different potentials in Fig.~\ref{fig:singularity}. To present our numerical data, we introduce
\begin{equation}\label{defR2}
  R_1=z\psi', \quad R_2=\ln\left(\frac{z_H^2}{z^2}-h\right), \quad R_3=4z^{2-d}e^{\chi/2}A_t'\,,
\end{equation}
with $h=e^{-\chi/2}f/z^{1+d}$. For the Kasner solution, one has $ \lim_{z\rightarrow\infty} R_i(z)=\mathrm{const.}$ with $i=1,2,3$. When the kinetic term dominates the dynamics, the numerical behavior satisfies the universal asymptotic form~\eqref{asyfchiaphi} [Fig.~\ref{fig:singularity}(b)]. In contrast, when the condition is not satisfied, in particular, Eq.~\eqref{constV1} is violated, numerical results exhibit behaviors that are quite distinct from the Kasner form. For potentials with exponential forms, we observe behaviors with strong oscillations all the way down to the singularity [Figs.~\ref{fig:singularity}(c) and~\ref{fig:singularity}(d)]. Numerical details and more examples are provided in Appendix~\ref{SLsingularity}. 

Before concluding this section, it is worth discussing the interior geometry a bit more with a non-trivial potential $V(|\Psi|^2)$. In the simplest holographic superconductor described by a free charged scalar in planar AdS black holes, it was found in Ref.~\cite{Hartnoll:2020fhc} that below the critical temperature $T_c$ there are intricate dynamical behaviors before ending at a spacelike Kasner singularity, including collapse of the Einstein-Rosen bridge, Josephson oscillations in the condensate and possible Kasner inversions. It would be interesting to see how much of these epochs could persist with non-trivial potential as compared to the free scalar case~\cite{Hartnoll:2020fhc}. 
We are not able to solve the problem thoroughly due to the complicated forms of Eqs.~\eqref{smeompsi}-\eqref{smeomf}. However, note that such intricate dynamics happens when the scalar field is quite small for which the non-linear terms of $V$ is not important. Therefore, one anticipates that the collapse of the Einstein-Rosen bridge and Josephson oscillations are not sensitive to the potential $V$. It was confirmed by checking some cases explicitly, see Fig.~\ref{fig:oscil1} and more discussions in Appendix~\ref{Vdynamics}. On the other hand, when the value of $\psi$ becomes large (for example, the case with temperature far below $T_c$ in Ref.~\cite{Hartnoll:2020fhc}), the interactions of the scalar field become important and the collapse of the Einstein-Rosen bridge and subsequent Josephson oscillations derivate from the free scalar case. We also stress that when the condition~\eqref{constV1} is broken, the asymptotic behavior around the spacetime singularity will be different from the Kasner behavior and the Kasner inversion is also broken.

\begin{figure}
\centering
  \includegraphics[width=1.0\textwidth]{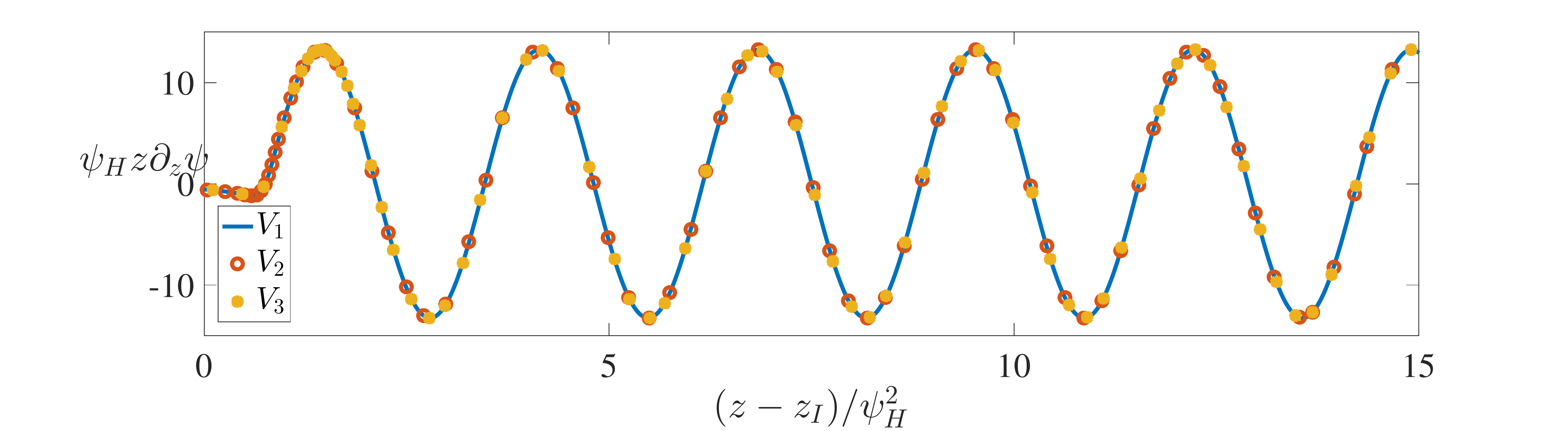}
  \caption{Behaviors of $z\psi'$ around the ``would-be" inner horizon $z=z_I$ that is destroyed by the condensate of the charged scalar field $\psi$. We have considered three different potentials $V_1(x)=x-6, V_2(x)=x+2x^3-6$ and $V_3(x)=-99x+\sinh(100x)-6$ with $\psi_H=\psi(z_H)=10^{-3}$ and $Z=1$.}
  \label{fig:oscil1}
\end{figure}
%

\section{Discussion}
We have shown a no inner-horizon theorem for black holes with charged scalar hair and discussed the possible asymptotic geometries near the singularity. Note that to prove the theorem we do not make direct use of the scalar potential $V$, thus our result applies to quite generic spacetimes. Since Eqs.~\eqref{minusQ} and~\eqref{km1Q} do not depend on $A_t$, our no inner-horizon results also apply to the Einstein-scalar theory in absence of $A_\mu$.  We then discussed the geometry near the singularity. The Kasner form of solutions has been obtained when the potential terms can be neglected. However, we have found strong numerical evidence for the existence of novel oscillating behavior all the way down to the singularity when the potential terms become important to the geometry.

Although one could engineer complicated hairy black holes by adjusting the scalar potential, our results suggest that the inner structure behind hairy black holes is pretty simple. There is no way to construct spherical and planar black holes with a Cauchy horizon for both Einstein-scalar (see \emph{e.g.} Ref.~\cite{Bronnikov:2001tv} for spherical case) and Einstein-Maxwell-charged scalar theories. Our theorem has some direct significance for the SCC by showing that the Cauchy horizon in large classes of theories will be definitely removed by a non-trivial scalar hairy. For example, it has been recently argued that SCC can be violated under the linear charged scalar perturbation of RN black holes in dS space~\cite{Cardoso:2018nvb,Mo:2018nnu,Dias:2018ufh}. Nevertheless, this linear perturbation result breaks down by considering non-linear effect of the charged scalar field.  As the energy accumulates near the inner horizon, non-linear effects from the scalar field cannot be neglected. Furthermore, if one considers time-dependence, the static hairy black hole could be the final state of the dynamically formed black hole. Notice that SCC asks for the instability and ensuing disappearance of Cauchy horizons. We have proved in this paper that the backreaction of the charged scalar field can sufficiently modify the background geometry and the resulted hairy black hole, if exists, has no Cauchy horizon\,\footnote{The full non-linear evolution of the massless charged scalar field in dS space was studied in Ref.~\cite{Zhang:2019nye}. It found that the non-linear effect can restore the SCC in the sense that the black hole is driven away from the initial highly near-extremal background by the charged perturbation.}.

In the present study, we have limited ourselves to black holes with maximally symmetric horizon, it would be interesting to consider more general cases with inhomogeneous spacetimes and also with additional forms of matter. We have explicitly shown that some hyperbolic black holes can have an inner horizon with charged timelike singularity (see Fig.~\ref{fig:hyperbolic}). It is interesting to further understand the interior of hyperbolic case. The dynamics near singularity seems to allow different behaviors from the Kasner form~\eqref{kasner}. It is desirable to understand this feature in the future.

\textbf{Note added}.--- While this work was being completed, the work~\cite{Hartnoll:2020fhc}  appeared in arXiv, which discusses the interior dynamics for planar black holes by considering a free charged scalar. Similar construction was used to prove the no Cauchy horizon feature for the planar horizon topology.

\section*{Acknowledgements}
We thank S.~A.~Hartnoll for helpful conversation. This work was partially supported by the National Natural Science Foundation of China Grants No.12075298, No.11991052, No.12047503, No.11690022, No.11821505 and No.11851302, by the Key Research Program of Frontier Sciences of CAS and by the National Key Research and Development Program of China Grant No.2020YFC2201501.

\appendix

\section{Example for Hyperbolic Black Holes with Inner-Horizon}\label{HBexample}
For the hyperbolic black holes, our method adopted in the main text can not be appropriate to prove the black hole no-inner horizon or that the inner horizon of black hole with charged scalar hairs may exist. In this section, we show that for the hyperbolic case ($k=-1$) it is indeed possible to have an inner horizon.

In order to have much higher numerical stability and to improve the error control, we further recombine the functions into the following form:
\begin{equation}\label{neweqs1}
  h:=fe^{-\chi/2}z^{-1-d},\quad \tilde{Q}:=-\frac{2\kappa_N^2}{\omega_{(d)}}Q=z^{2-d}e^{\chi/2}A_t'\,.
\end{equation}
By using these new variables, we can rewrite Eqs.~\eqref{smeompsi}-\eqref{smeomf} into the following form and then numerically solve the variables $\{\chi,h, \psi, \tilde{Q}, A_t\}$.
\begin{align}
\chi' =& \frac2{d}\left[\frac{\psi^2A_t^2q^2}{h^2z^{2d+1}}+z\psi'^2\right]\,,\label{eqschi0}\\
h'=&-\frac{(d-1)k}{z^d}e^{-\chi/2}+\frac{e^{-\chi/2}}{d}\left(\frac{\tilde{Q}^2z^{d-2}}{2}+\frac{V(\psi^2)}{z^{d+2}}\right)\,,\label{eqsf0}\\
\tilde{Q}'=&\frac{2\psi^2A_tq^2}{z^{2d+1}h},\qquad A_t'=\tilde{Q}e^{-\chi/2}z^{d-2}\,,\label{eqsA0}\\
\psi''=& -\left(\frac{h'}{h}+\frac{1}z\right)\psi'+\left(\frac{\dot{V}(\psi^2)e^{-\chi/2}}{z^{d+3}h}-\frac{A_t^2q^2}{h^2z^{2d+2}}\right)\psi\,,\label{eqspsi0}
\end{align}
where we have set $Z=1$.

We stress that a Cauchy horizon for the hyperbolic black holes exists only for some specific value of the parameters. We now present a concrete numerical example for the hyperbolic black hole with an inner Cauchy horizon. The four dimensional model ($d=2$) is given by
\begin{equation}
V(\psi^2)=-6+m^2\psi^2\,, \quad Z=1\,,
\end{equation}
with $m^2=-0.18388$ and $q=1.5$. At the event horizon $z_H=1.193936$, we choose the following initial conditions:
\begin{equation}\label{soluq1}
\begin{split}
&\chi(z_H)=h(z_H)=A_t(z_H)=0\,,\\
& \psi(z_H)\approx1.10683410,\quad \psi'(z_H)\approx0.115816263,\quad \tilde{Q}(z_H)\approx0.650999915\,.
  \end{split}
\end{equation}
\begin{figure}[h!]
\begin{center}
\includegraphics[width=0.45\textwidth]{fig/full1.pdf}
\includegraphics[width=0.45\textwidth]{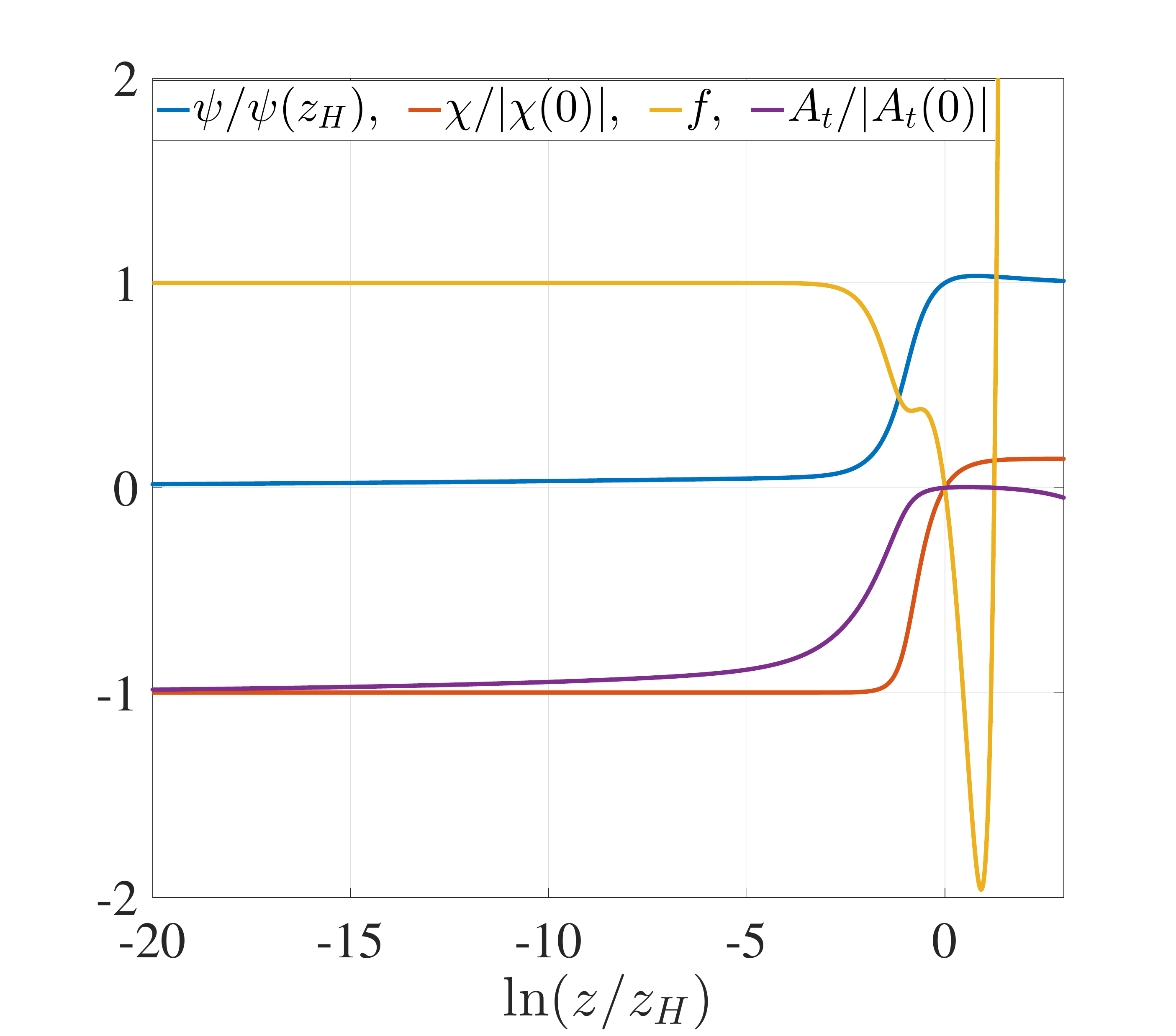}
\caption{Numerical solution for hyperbolic case ($k=-1$) with the boundary conditions~\eqref{soluq1}. The hairy black hole has the event horizon at $z_H=1.193936$. There is a Cauchy horizon at $z_I\approx4.15699837$ and all functions are smooth at two horizons.  From the right panel, we see that $f(0)=1, \psi(0)=0$, $\chi(0)$ and $A_t(0)$ are both finite, which implies that this solution is indeed an asymptotically AdS black hole. We have considered the four dimensional model with $V(\psi^2)=-6-0.18388\psi^2, Z=1$ and $q=1.5$. }
\label{fig:full1}
\end{center}
\end{figure}

We then integrate Eqs.~\eqref{eqschi0}-\eqref{eqspsi0} with $k=-1$ numerically from $z=z_H+\varepsilon_1$ to $z\rightarrow\infty$ (here we set $\varepsilon_1=10^{-9}$). For example, we have used the solver \verb"ode45" of MATLAB with accuracy control ``\verb"odeset('RelTol',1e-13,'AbsTol',1e-13)"''. As Eqs.~\eqref{eqschi0}-\eqref{eqspsi0} will meet coordinate singularity at $z=z_I$, numerical solver will fail at this singularity and cannot directly pass through this point. To overcome this issue, we slightly modify Eqs.~\eqref{eqschi0},\eqref{eqsA0} and \eqref{eqspsi0} into the following form:
\begin{equation}\label{modifpsi1}
\begin{split}
\chi' = &\frac2{d}\left[\frac{\psi^2A_t^2q^2}{(h+i\varepsilon_2)^2z^{2d+1}}+z\psi'^2\right],\qquad\quad \tilde{Q}'=\frac{2\psi^2A_tq^2}{z^{2d+1}(h+i\varepsilon_2)}\,,\\
  \psi''=& -\left(\frac{h'}{h+i\varepsilon_2}+\frac{1}z\right)\psi'+\left(\frac{\dot{V}(\psi^2)e^{-\chi/2}}{z^{d+3}(h+i\varepsilon_2)} -\frac{A_t^2q^2}{(h+i\varepsilon_2)^2z^{2d+2}}\right)\psi\,.
  \end{split}
\end{equation}
This leads to the fact that the solutions all have imaginary parts with order $\mathcal{O}(\varepsilon_2)$. Then we carefully tune $\varepsilon_2$ from $10^{-3}$ to $10^{-9}$ to verify the convergency of real parts for variables $\{h(z),\psi(z),A_t(z),\chi(z),\tilde{Q}(z), \psi'(z)\}$. For example, when $\varepsilon_2=10^{-5}, 10^{-7}, 10^{-9}$, we have
\begin{equation}
\psi(6)\approx1.13489227 + 3\times10^{-6}i,~~1.13540830+6\times10^{-7}i,~~1.13541591+1.0\times10^{-8}i\,,
\end{equation}
respectively. Thus, the real parts will convergent and we can ignore all imaginary parts when $\varepsilon$ is small enough.  The numerical results are shown in Fig.~\ref{fig:full1}.
It is clear that the inner horizon appears at $z_I\approx4.15699837$ at which $f(z_I)=A_t(z_I)=0$. Note that if one wants to obtain higher accuracy, \emph{e.g.} by setting $\varepsilon<10^{-9}$, the accuracy of initial conditions~\eqref{soluq1} should also be improved. Notice also that the hairy black hole is asymptotically AdS as $z\to 0$, see the right panel of Fig.~\ref{fig:full1}.

\section{Properties of Gauge Sector}\label{gauge}
The properties of gauge potential $A_t$ play a crucial role in  understanding the structure of interior geometry behind the event horizon. In this section we will discuss some general features for the gauge sector. In particular, we will present a few lemmas of the gauge sector, which are important to build our no inner-horizon theorem in the main text.

~\\
\textbf{Lemma 1: } \textit{For hairy charged black hole (i.e. $\psi$ and $A_t$ are not zero somewhere), $A_t(z_i)$ must be zero at any horizon $z=z_i$ where $f(z_i)$ vanishes.}\\

\textit{Proof of \textbf{Lemma 1}:} This proof uses the smoothness of the spacetime geometry away from the singularity. When $f(z_i)=0$, to insure the smoothness around $z=z_i$ for other functions, we have the following two choices:
\begin{center}
(a) $A_t(z_i)=0$, \qquad or\qquad (b) $A_t(z_i)\neq0$,\;\;$\psi(z_i)=0$\,.
\end{center}
What we need is to exclude the second case.
We first assume that case (b) is true. As $\psi(z)$ is smooth and nonzero somewhere, we have the following Taylor expansion:
\begin{equation}\label{taypsi1}
  \psi=\psi_n\delta^n+\psi_{n+1}\delta^{n+1}+\cdots,~~n\geq1\,,
\end{equation}
with $\psi_n\neq0$ and $\delta=z-z_i$. Similarly, $f$ has the expansion
\begin{equation}\label{tayf1}
  f=f_l\delta^l+f_{l+1}\delta^{l+1}+\cdots,~~l\geq1\,.
\end{equation}
with $f_l\neq0$. Since we have assumed $A_t(z_i)\neq0$, smoothness of Eq.~\eqref{smeomchi} implies $n\geq l$. Up to the leading order, Eq.~\eqref{smeompsi} becomes
\begin{equation}\label{eqfornl1b}
  \psi_n n(n-1)\delta^{n-2}=-n l\psi_n\delta^{n-2}-\frac{e^{\chi(z_i)}A_t^2(z_i)q^2\psi_n}{f_l^2}\delta^{n-2l}\,,
\end{equation}
where we have used $Z(0)=1$. Then we find that
\begin{equation}\label{eqfornl1}
  -e^{\chi(z_i)}A_t^2(z_i)q^2=f_l^2n(n-1+l)\delta^{2l-2}\,.
\end{equation}
This is impossible because the left side is negative, while the right side is not negative. Therefore, we arrive at the result of Lemma 1. $\square$

~\\
\textbf{Lemma 2:  }\textit{In an interval $(z_1,z_2)$, if $f(z)\geq0$, then $A_t(z)^2$ has no local maximum.}\\

\textit{Proof of \textbf{Lemma 2}: }We only need to consider the case for which $A_t$ is not a constant in the interval $(z_1,z_2)$. We write Eq.~\eqref{smeomAt} into the following form
\begin{equation}\label{at2s1}
  (A_t^2)''-\left(\frac{d-2}z-\frac12\chi'+\frac{\mathrm{d} \ln Z(\psi^2)}{\mathrm{d} z}\right)(A_t^2)'=\frac{4\psi^2A_t^2q^2}{z^2fZ(\psi^2)}+2A_t'^2\,.
\end{equation}
Then in the interval $(z_1,z_2)$ with $f>0$, we see that
\begin{equation}\label{at2s2}
  (A_t^2)''-\left(\frac{d-2}z-\frac12\chi'+\frac{\mathrm{d} \ln Z(\psi^2)}{\mathrm{d} z}\right)(A_t^2)'\geq0\,.
\end{equation}
If there is a local maximum which locates at $z_m\in(z_1,z_2)$, we have the following Taylor expansion
\begin{equation}\label{maxzmA2}
  A_t^2=A_0-A_s(z-z_m)^{2s}+\cdots\,,
\end{equation}
with $A_s>0$ and the integer $s\geq1$. Taking it into Eq.~\eqref{at2s2}, we find
\begin{equation}\label{at2s3}
  -2s(2s-1)A_s(z-z_m)^{2s-2}+\mathcal{O}((z-z_m)^{2s-1})\geq0\,.
\end{equation}
As the left side of Eq.~\eqref{at2s3} is negative, Eq.~\eqref{at2s3} cannot be satisfied. Thus, there is no local maximum in the interval $(z_1,z_2)$. In addition, if $A_t(z)$ is continuous at two endpoints, then the maximum of $A_t(z)^2$ in the interval $[z_1,z_2]$ is given by $\max\{A_t(z_1)^2, A_t(z_2)^2\}$. $\square$

~\\

The charge degrees of freedom behind the surface generate a non-zero electric flux and can be  characterized by
\begin{equation}
Q(z)=\frac{1}{2\kappa_N^2}\int_{\Sigma} Z(|\Psi|^2)\, {}^\star F=-\frac{\omega_{(d)}}{2\kappa_N^2}Z(\psi^2)z^{2-d}e^{\chi/2}A_t'\,,
\end{equation}
where $\omega_{(d)}$ is the volume of the section with $t$ and $z$ fixed and we have used the ansatz~\eqref{ansatz}. We have the following Lemma for $Q(z)$.\\

\textbf{Lemma 3: }\textit{In an interval $(z_1,z_2)$ with $f(z)\geq0$ and $A_t(z_1)=0$, $Q(z)^2$ is monotonic increasing in the interval $(z_1,z_2)$.}\\

\begin{figure}[h]
\centering
  \includegraphics[width=0.7\textwidth]{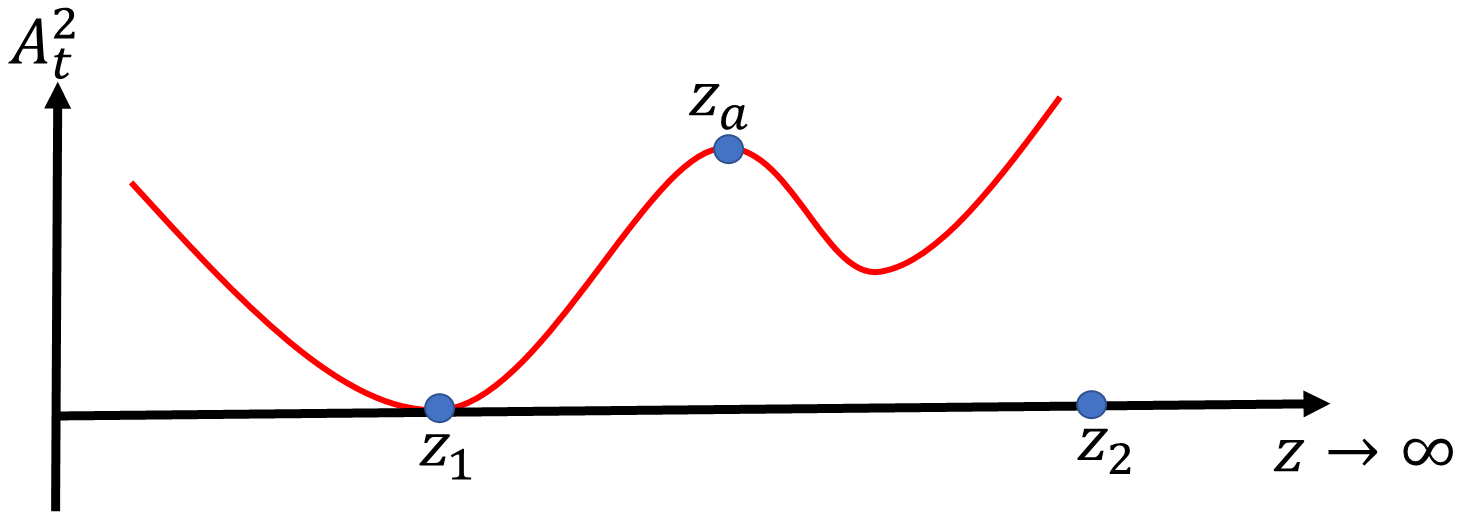}
  \caption{As $A_t(z_1)^2=0$, $A_t(z)^2$ increases at beginning when $z>z_1$. If $A_t(z)^2$ begins to decrease after $z>z_a$, $A_t(z_a)^2$ will be a local maximum. However, there should be no any local maximum according to \textbf{Lemma 2}. Thus, the point $z_a$ should not exist.} \label{figAt1}
\end{figure}

\textit{Proof of \textbf{Lemma 3}: } We first rewrite Eq.~\eqref{smeomAt} as
\begin{equation}\label{neeqsA1}
  (z^{2-d}e^{\chi/2}Z A_t')'=\frac{2e^{\chi/2}\psi^2q^2}{z^df}A_t\,.
\end{equation}
Making use of Eq.~\eqref{smeomAt}, we then have
\begin{equation}\label{neeqsA1}
  Q'=-\frac{\omega_{(d)}}{2\kappa_N^2}\frac{2e^{\chi/2}\psi^2q^2}{z^df}A_t\Rightarrow QQ'=-Q\frac{\omega_{(d)}}{2\kappa_N^2}\frac{2e^{\chi/2}\psi^2q^2}{z^df}A_t=\frac{\omega_{(d)}^2Zz^2e^{\chi}\psi^2q^2}{2\kappa_N^4 z^{2d}f}A_tA_t'\,.
\end{equation}
Thus, we obtain
\begin{equation}\label{neeqsA2}
  (Q^2)'=\frac{\omega_{(d)}^2Zz^2e^{\chi}\psi^2q^2}{2\kappa_N^4 z^{2d}f}(A_t^2)'\,.
\end{equation}
From \textbf{Lemma 2}, in the interval $(z_1,z_2)$ with $f(z)\geqslant 0$, $A_t^2$ has no local maximum. Since $A_t(z_1)=0$, one obtains $(A_t^2)'\geq0$ (see Fig.~\ref{figAt1} for a schematic explanation). Then, we immediately have
\begin{equation}
  (Q^2)'\geqslant 0\,,
\end{equation}
in an interval $(z_1,z_2)$ with $f(z)\geq0$ and $A_t(z_1)=0$. Note that when $f=0$, Eq.~\eqref{neeqsA1} seems to be singular. Nevertheless, the smoothness of equations of motion~\eqref{smeompsi}-\eqref{smeomf} insures that this is a removable singularity. Thus the desired result follows.

One interesting feature is that the timelike singularity must be charged. The hairless solution is described by the Reissner-Nordstr\"{o}m black hole, for which there is no any medium to conduct the charge, thus the charge has to stay at the original point. When the charged scalar hair develops, the charge can escape from the singularity via the complex scalar field. Therefore, one could ask whether the original point is neutral or not after the charged scalar condenses. The answer is negative. The proof is as follows. As the singularity is timelike, there must be a horizon at $z=z_s$ and $f(z)>0$ when $z>z_s$. Then, we obtain from~ \textbf{Lemma 3} that
\begin{equation}\label{chargsing1}
  Q(z\to \infty)^2>Q(z_s)^2\geq0\,,
\end{equation}
where we have assumed that the spacetime singularity appears as $z\to\infty$.
Thus, the timelike singularity must be charged, suggesting that near the timelike singularity, the gravitational attraction always dominates the electric repulsion.

\section{Asymptotic Solutions Near the Spacelike Singularity}\label{SLsingularity}
As we have shown in the main text that for a large class of hairy black holes there exists no inner horizon and the spacetime ends at a spacelike singularity. In this section, we present the analytical analyzes as well as numerical details for constructing the geometry near the spacelike singularity. To simplify our discussion, we shall specify $Z=1$ so that the theory becomes a standard Einstein-Maxwell-scalar theory.

We begin our discussion with the case that the potential of scalar field can be neglected near the singularity. We will show that in this case the spacelike singularity always has the asymptotic Kasner geometry. We then discuss what will happen if this condition is broken.

\subsection{Analytical analyze}

Near the singularity, the functions will be divergent, which challenges the numerical precision. So we first give some analytic discussion about the geometry near the singularity.

When the contribution from the scalar effective potential can be neglected, we obtain the following equations from Eqs.~\eqref{eqschi0}-\eqref{eqspsi0} near the singularity as $z\rightarrow\infty$:
\begin{align}
\chi' =&\frac2{d}\left[\frac{\psi^2A_t^2q^2}{h^2z^{2d+1}}+z\psi'^2\right],\label{eqschi}\\
h'=&\frac{\tilde{Q}^2z^{d-2}e^{-\chi/2}}{2d},\label{eqsf}\\
\tilde{Q}'=&\frac{2\psi^2A_tq^2}{z^{2d+1}h},\qquad A_t'=\tilde{Q}e^{-\chi/2}z^{d-2},\label{eqsA}\\
\psi''=&- \left(\frac{h'}{h}+\frac{1}z\right)\psi'-\frac{A_t^2q^2}{h^2z^{2d+2}}\psi.\label{eqspsi}
\end{align}
It is obvious from Eq.~\eqref{eqsf} that $h(z)$ is a monotonic increasing function when $z$ is large enough. Since we have assumed that the singularity is spacelike, we have $h<0$ as well as
\begin{equation}
\lim_{z\rightarrow\infty}h(z)=-h_0<0\,.
\end{equation}
Then, we find $\mathcal{O}(h')<\mathcal{O}(1/z)$. Therefore, we obtain from Eq.~\eqref{eqsA} that
\begin{equation}\label{boundat1}
  A_t'^2=2dh'e^{-\chi/2}z^{d-2}<\mathcal{O}(z^{d-3})\,,
\end{equation}
and thus $|A_t|<|\mathcal{O}(z^{(d-1)/2})|$. Here we have used the fact that $e^{-\chi/2}\leq\mathcal{O}(z^0)$ since $\chi$ is a monotonic increasing function and is finite at the event horizon. This leads to that the coefficient of the last term of Eq.~\eqref{eqspsi} satisfies
\begin{equation}
\frac{A_t^2q^2}{h^2z^{2d+2}}<\mathcal{O}(1/z^{d+3})<\mathcal{O}(1/z^{2})\,.
\end{equation}
Therefore, the last term of Eq.~\eqref{eqspsi} can be neglected.
The solutions of $\psi$ in large $z$ limit then reads
\begin{equation}\label{solupsi1}
  \psi(z)\sim \sqrt{d}\alpha\ln z\,,
\end{equation}
with $\alpha$ a constant. Taking it into Eq.~\eqref{eqschi}, we can find $\chi'=2\alpha^2/z$. Finally, we obtain a simple result for the geometry near the singularity.
\begin{equation}\label{asyfchiaphi2}
\begin{split}
\psi=\sqrt{d}\alpha\ln z+\cdots, \quad A_t'=E_{s} z^{d-2-\alpha^2}+\cdots\,,\\
\chi=2\alpha^2\ln z+\cdots,\quad  f=-f_{s}z^{1+d+\alpha^2}+\cdots\,,
\end{split}
\end{equation}
where $E_s$ and $f_s$ are constants.

Changing the $z$ coordinate to the proper time $\tau$ via $\tau\sim z^{-(1+d+\alpha^2)/2}$, we obtain
\begin{equation}\label{smkasner}
ds^2=-d\tau^2+c_t \tau^{2 p_t} d t^2+c_{s} \tau^{2 p_{s}}d\Sigma^2_{d,k},\quad  \psi(z)=-p_{\psi} \ln \tau\,,
\end{equation}
with
\begin{equation}
p_t=\frac{1-d+\alpha^2}{1+d+\alpha^2}, \quad p_s=\frac{2}{1+d+\alpha^2}, \quad p_{\psi}=\frac{2\sqrt{d}\alpha}{1+d+\alpha^2}\,.
\end{equation}
One can immediately check that
\begin{equation}
p_t+d p_s=1,\quad p_{t}^2+d p_{s}^2+p_{\psi}^2=1\,,
\end{equation}
and therefore the geometry around the spacelike singularity has the Kasner form.
We now arrive at the conclusion: when the scalar potential can be neglected in the spacelike singularity, the asymptotic solutions are of Kasner type in Eq.~\eqref{smkasner}.

Note that in above discussion we have assumed that the scalar kinetic term should dominate the dynamics near the singularity. One should at least has the following constraint:
\begin{equation}\label{constV1b}
  \lim_{z\rightarrow\infty}\frac{|V(\psi^2)|}{z^{d+1+\alpha^2}}\ll 1\,.
\end{equation}

In particular, it allows the scalar potential $V$ to be arbitrary algebraic functions, including polynomial functions. However, if one chooses a potential that diverges exponentially or even worse, the condition~\eqref{constV1b} will be broken and we can not obtain~\eqref{asyfchiaphi2}. For example, we take
\begin{equation}\label{exampv1}
  V(\psi^2)=P(\psi^2)+\sinh(\gamma\,\psi^2)\,,
\end{equation}
with $P(x)$ a polynomial. When $\gamma=0$, we expect to obtain~\eqref{asyfchiaphi2} no matter how high order of polynomial one considers. In contrast, once we choose $\gamma>0$, the asymptotic solution will be different from~\eqref{asyfchiaphi2}. This can be understood as follows. Suppose we are in the Kasner regime, where we have $\psi\sim \sqrt{d}\alpha\ln z$ at large $z$. After including the second term of Eq.~\eqref{exampv1}, one has
\begin{equation}\label{break}
\frac{|V(\psi^2)|}{z^{d+1+\alpha^2}}\sim \frac{e^{d\gamma \alpha^2(\ln z)^2}}{z^{d+1+\alpha^2}}>\frac{e^{d\gamma \alpha^2 (\kappa \ln z)}}{z^{d+1+\alpha^2}}=\frac{z^{\kappa\gamma d\alpha^2}}{z^{d+1+\alpha^2}}\,.
\end{equation}
Here $\kappa$ is a constant for which we only demand $\kappa<\ln z$. Therefore, $\kappa$ can be sufficiently large as we approach the singularity. In particular, when $\kappa>\frac{d+1+\alpha^2}{\gamma d \alpha^2}$, the numerator of Eq.~\eqref{break} is larger than the denominator and therefore the constraint~\eqref{constV1b} is broken. As a consequence, no matter how small the value of $\gamma$ is, the Kasner solution is expected to be modified when $z>z_c$, where the critical point is given by
\begin{equation}\label{zc}
z_c=c_0\, e^{\frac{d+1+\alpha^2}{\gamma d \alpha^2}}\,,
\end{equation}
where $c_0$ is a coefficient that depends on a model one considers. To be self-consistent, we need $z_c\gg z_H$ such that the solutions are approximately in the Kasner regime when $\gamma=0$.

\subsection{Numerical check}

In this section we examine the geometry inside hairy charged black holes by numerically solving Eqs.~\eqref{eqschi0}-\eqref{eqspsi0}. We are particularly interested in the behavior near the singularity. We also would like to check our analytic results obtained in the last part.

Note that the Kasner type solution means $\{h, Q, A_t, z\chi', z\psi'\}$ approach to constant. To present our numerical data, we further introduce
\begin{equation}
  R_1=z\psi',~R_2=\ln\left(\frac{z_H^2}{z^2}-h\right),~~R_3=4\tilde{Q}\,,
\end{equation}
for which the Kasner solution yields
\begin{equation}
  \lim_{z\rightarrow\infty} R_i(z)=\mathrm{const.}, \quad i=1,2,3\,.
\end{equation}

\begin{figure}[h]
\centering
  \includegraphics[width=.45\textwidth]{fig/case1.pdf}\qquad
  \includegraphics[width=.45\textwidth]{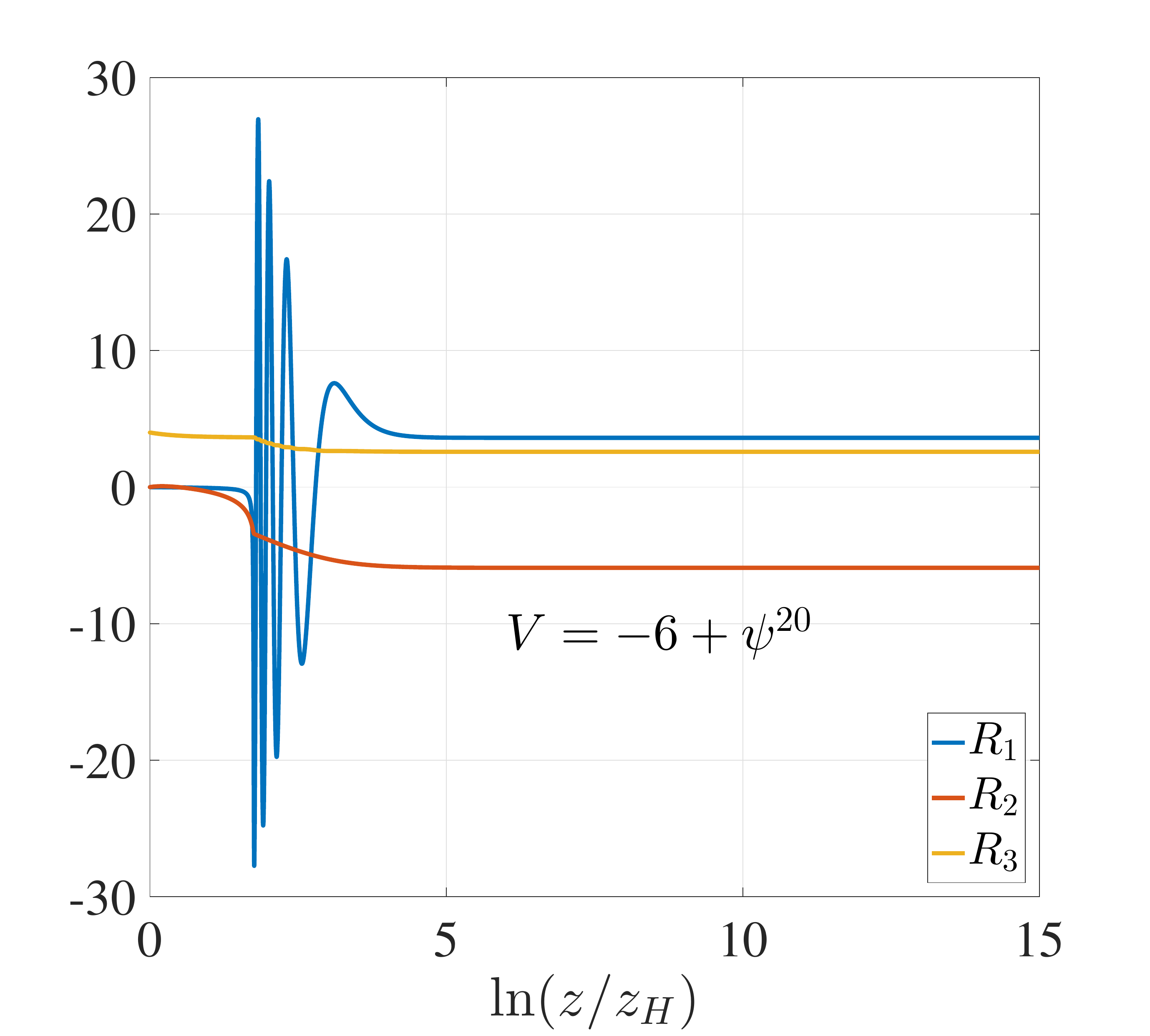}
  \caption{Numerical results for different polynomial scalar potentials. $\{R_1, R_2, R_3\}$ all approach to be constant  values as $z/z_H\rightarrow\infty$, confirming the Kasner form geometry near the singularity. We have fixed $Z=1$.}
 \label{figkasner1}
\end{figure}

We first consider the class of potentials
\begin{equation}\label{caseV1}
V(\psi^2)=-6+\psi^{2n}+\sinh(\gamma\psi^2)\,,
\end{equation}
and numerically solve Eqs.~\eqref{eqschi0}-\eqref{eqspsi0} by using the Runge-Kutta method.
We consider the planar black holes with $k=0$ and specify $d=2, z_H=1, q=1$. We choose $A_t'(z_H)=1, \chi(z_H)=0$ and $\psi(z_H)=1/2$ in all cases. As we only care about the inner geometry, we do not consider the UV completion for~\eqref{caseV1}. We first present the numerical result for the polynomial case, for which we set $\gamma=0$ and vary $n$. According to the discussion above, we anticipate Kasner type solutions~\eqref{asyfchiaphi2} as $z\to \infty$. In Fig.~\ref{figkasner1} we show our numerical results for $V=-6+\psi^{2}$ and $V=-6+\psi^{20}$. As expected, the functions $\{R_1, R_2, R_3\}$ all approach to constant when $z/z_H\rightarrow\infty$, confirming the Kasner form near the singularity.
\begin{figure}[h]
\centering
  \includegraphics[width=.45\textwidth]{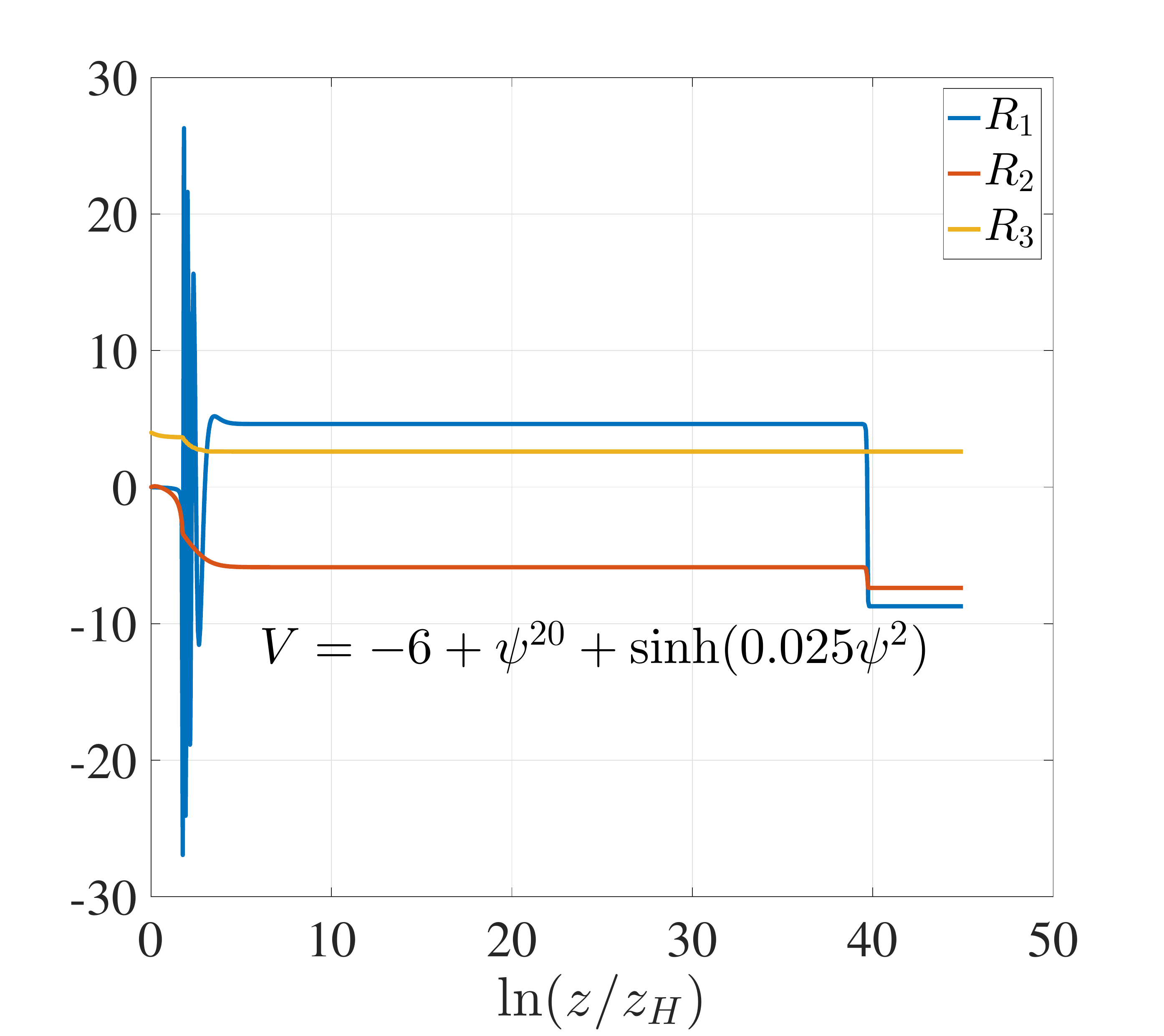}\qquad
  \includegraphics[width=.45\textwidth]{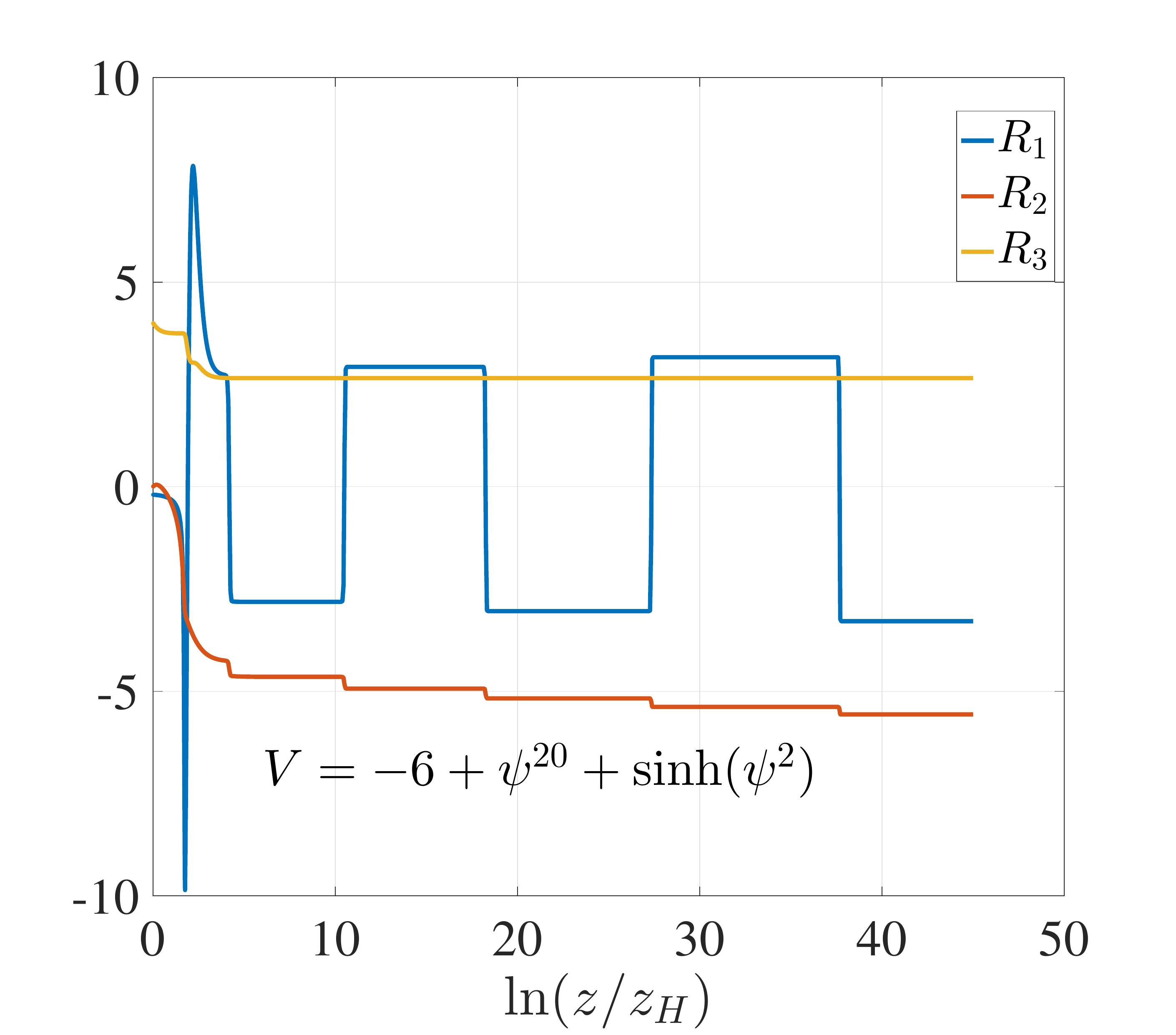}
  \caption{Numerical results for non-polynomial potentials. The behaviors near the singularity are different from the Kasner form. The oscillating behavior appears when $z$ is large. The observation of intensity oscillation becomes more and more manifest as $\gamma$ is increased.} \label{figkasner2a}
\end{figure}
\begin{figure}[h!]
  \includegraphics[width=.45\textwidth]{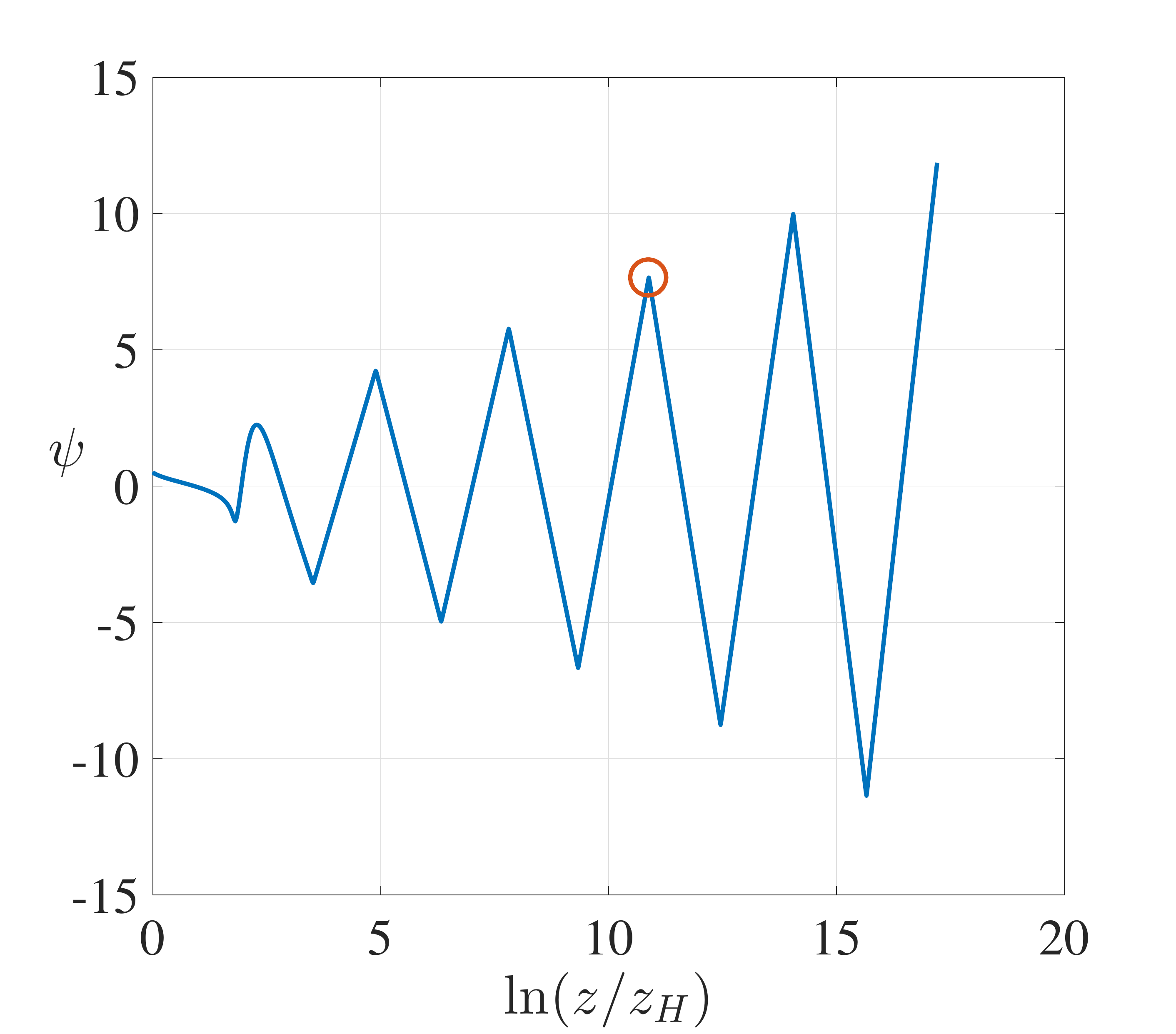}\qquad
  \includegraphics[width=.45\textwidth]{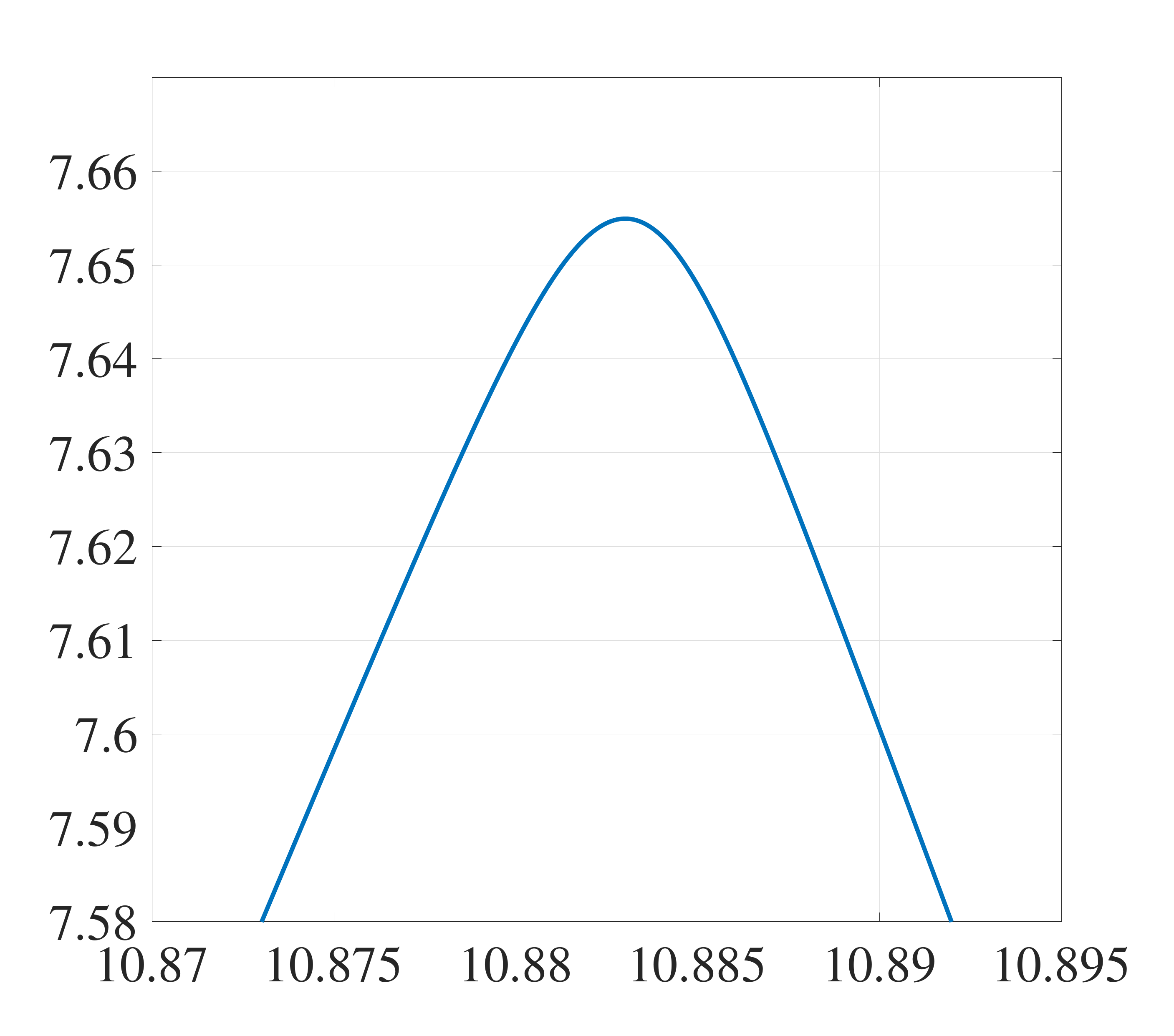}
  \caption{The behavior of the scalar field near the singularity for $V=-6+\psi^2+\sinh5\psi^2$. Left panel: $\psi$ as a function of $\ln(z/z_H)$. Right panel: we zoom in the peak in the red circle of the left panel, which shows that the solution is smooth near the peak.} \label{figcases2}
\end{figure}

As the value of $\gamma$ is increased, the contribution from $V$ to the geometry near the singularity becomes more and more important. The numerical results are presented in Fig.~\ref{figkasner2a}. For non-polynomial potentials which violate~\eqref{constV1b}, the asymptotic solutions are different from the Kasner behavior~\eqref{asyfchiaphi2}. We observe some strong oscillating behavior all the way down to the singularity. The oscillating behavior can be found more clearly from the behavior of $\psi$, which is shown in  Fig.~\ref{figcases2}.

We see that the scalar field does no longer satisfy the asymptotic behavior shown in Eq.~\eqref{asyfchiaphi2}. Interestingly, our numerical result implies that the asymptotic solution of $\psi$
approximately obeys
\begin{equation}\label{asyfchiaphi3}
\psi\sim \mathcal{T}_0(z)\,\mathcal{T}_1[\ln(z/z_H)]\,,
\end{equation}
where $\mathcal{T}_0(x)$ is a slow-varied function and $\mathcal{T}_1[\ln(z/z_H)]$ a periodic function of $\ln(z/z_H)$ (see the left panel of Fig.~\ref{figcases2}). We have carefully checked the convergence of our numerical results.

\begin{figure}
\centering
  \includegraphics[width=.45\textwidth]{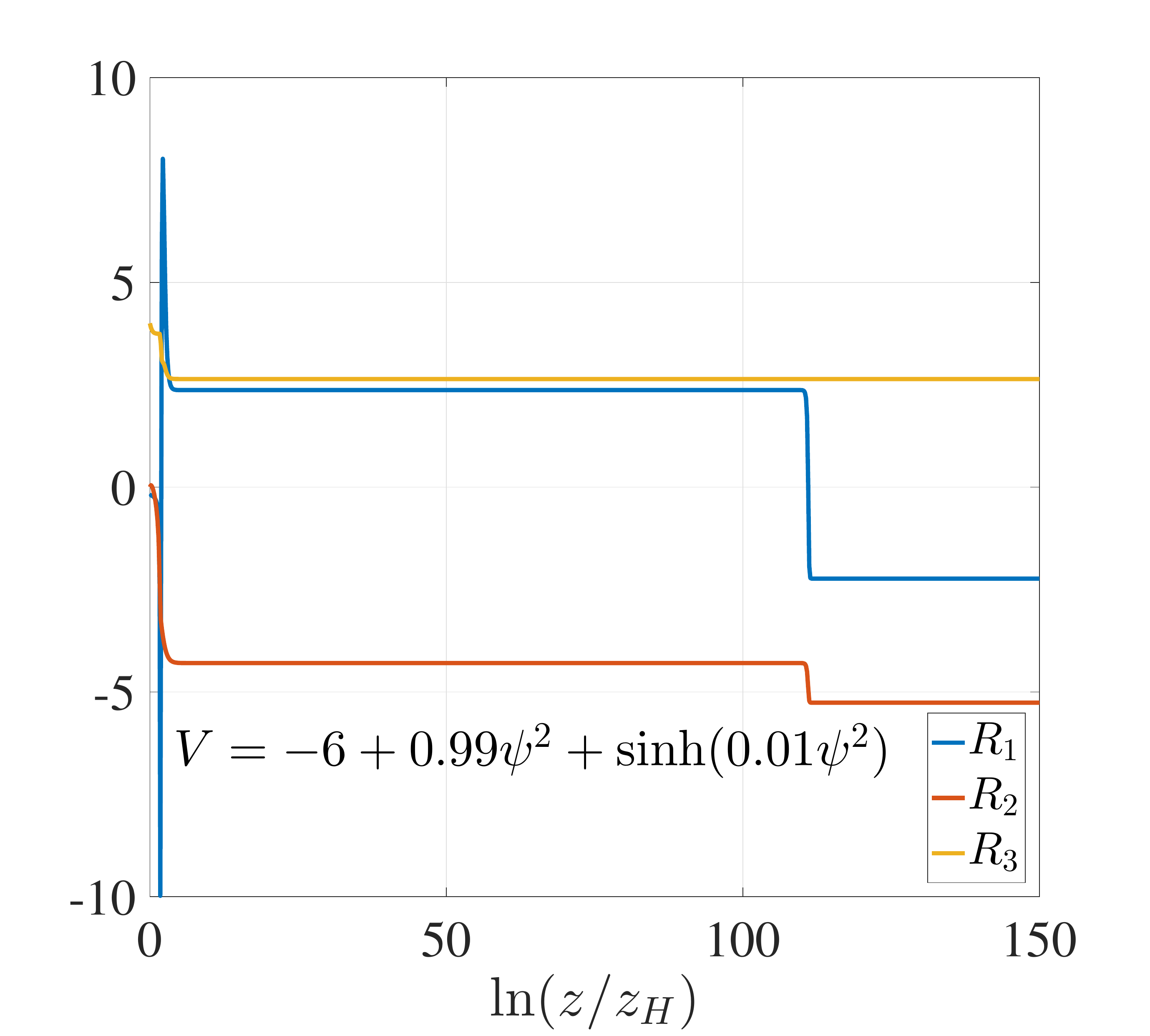}\qquad
  \includegraphics[width=.45\textwidth]{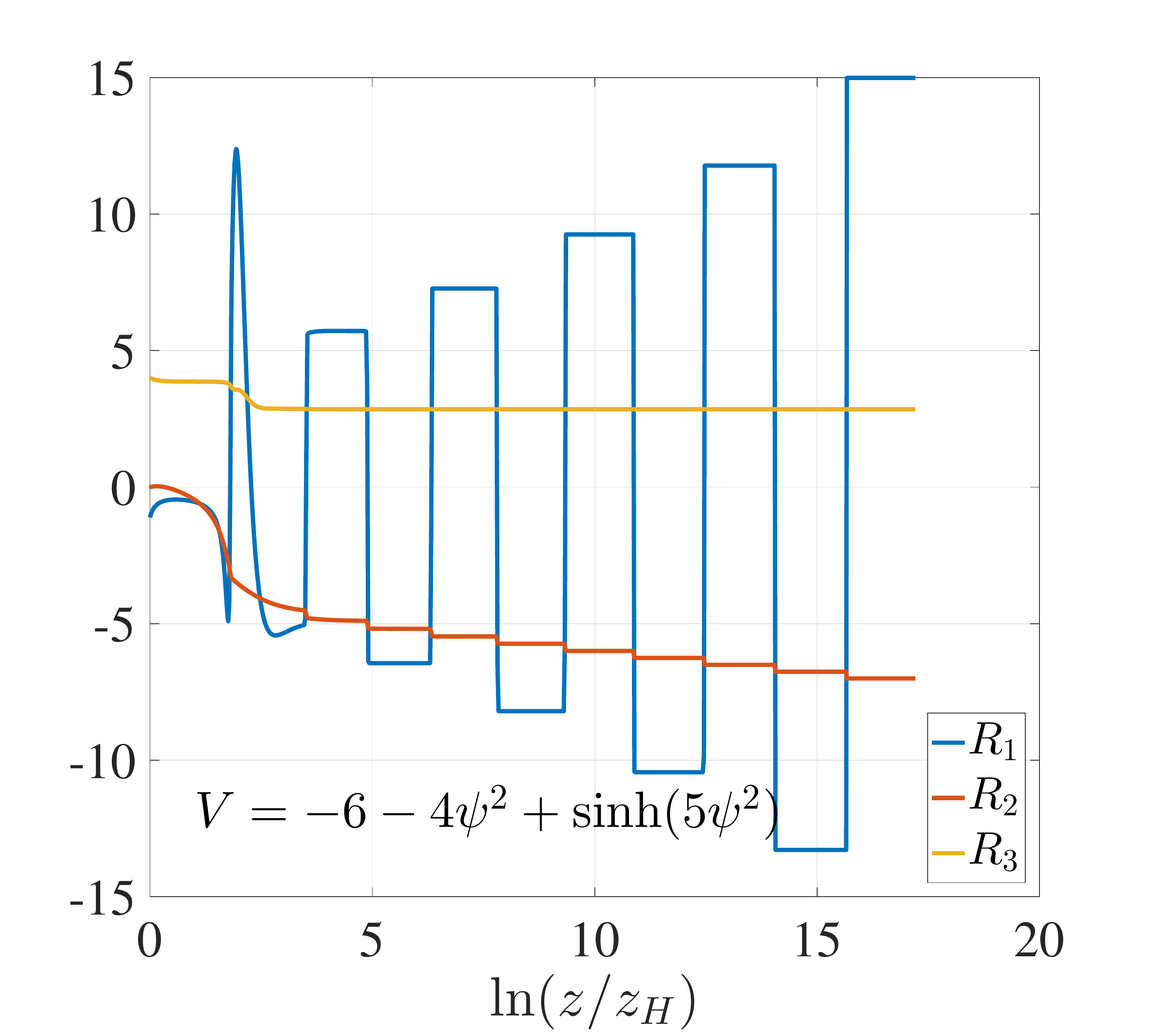}
  \caption{Numerical results for non-polynomial potentials which allow black holes in AdS. The behaviors near the singularity are different from the Kasner form. The oscillating behavior appears when $z$ is large. The observation of intensity oscillation becomes more and more manifest as $\gamma$ is increased.} \label{figkasner2}
\end{figure}

One may worry that for the choice of potentials in Eq.~\eqref{caseV1}, the asymptotic geometry near the black hole boundary is different for different $\gamma$. To avoid this issue, we consider the second class of potentials
\begin{equation}\label{potenvas1}
  V(\psi^2)=-6+(1-\gamma)\psi^2+\sinh(\gamma\psi^2)\,,
\end{equation}
for which the black holes are asymptotic AdS as $z\to 0$ and near the AdS boundary $V$ behaves as $V=-6+\psi^2+\cdots$. We obtain very similar oscillating behavior as the first class of potentials, see Fig.~\ref{figkasner2}.

\begin{figure}[h!]
  \includegraphics[width=.45\textwidth]{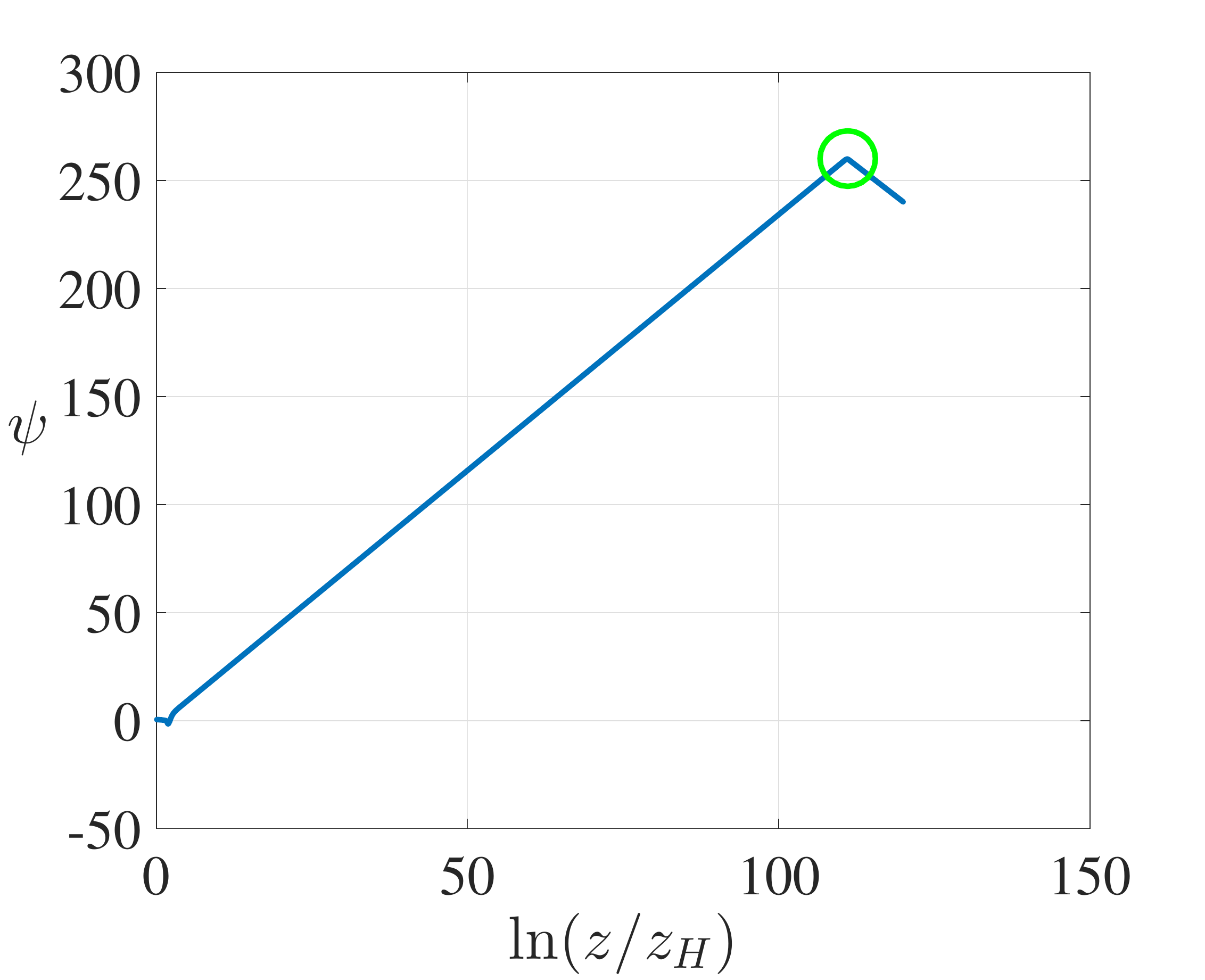}\qquad
  \includegraphics[width=.45\textwidth]{fig/zc2.pdf}
  \caption{Left panel: numerical results about the scalar field $\psi$ near the singularity for $V=-6+0.99\psi^2+\sinh(0.01\psi^2)$. We can still observe the logarithmic behavior $\psi\sim\ln(z/z_H)$ when $z<z_c\approx z_H\times10^{44}$ (green circle), which implies that the Kanser form solution is valid when $z<z_c$. Right panel: relationship between $z_c$ and $\gamma$, which obeys the approximated scaling behavior~\eqref{relzca1} and matches the theoretical prediction~\eqref{zc}.}
  \label{figcases3}
\end{figure}

For a potential of Eq.~\eqref{exampv1}, as we have discussed around Eq.~\eqref{break}, once $\gamma$ is turned on, no matter how small it is, the Kasner form~\eqref{asyfchiaphi2} would be broken. Numerically, it is not easy to verify this feature for small $\gamma$. This is because the exponential term will play role only when the scalar field is large enough, which means one has to solve the equations to sufficiently large $z$. Nevertheless, one anticipates from numerics that there exists a critical value $z_c$, beyond which the Kasner behavior would be modified. To check this point, we consider the class of potentials in Eq.~\eqref{potenvas1} and numerically study how $z_c$ depends on the parameter $\gamma$. We indeed observe expected features from our numerics shown in Fig.~\ref{figcases3}. In particular, from the left panel of Fig.~\ref{figcases3}, one finds that  there is a critical value of $z$ at which the Kasner solution ceases to be valid for a given small value of $\gamma$. We then numerically find the relationship between $z_c$ and $\gamma$, which is shown in the right panel of Fig.~\ref{figcases3}.

We obtain from Fig.~\ref{figcases3} that for small $\gamma$, there is a scaling behavior
\begin{equation}\label{relzca1}
  \gamma \ln(z_c/z_H)=\text{const.}\,,
\end{equation}
suggesting that the oscillating behavior will appear for large enough $z$, no matter how small $\gamma$ is. Interestingly, the scaling relation~\eqref{relzca1} is exactly we have obtained in Eq.~\eqref{zc}. We also compare the numerical results with our theoretical prediction~\eqref{zc} quantitatively. We obtain $\alpha$ by using the relation $\psi=\sqrt{d}\alpha\ln z$ when $z\lesssim z_c$ (the linear region in the left panel of Fig.~\ref{figcases3}). After fitting the coefficient $z_0$ in Eq.~\eqref{zc}, we find that the numerical results match with the theoretical prediction~\eqref{zc} quite well (see the solid line in the right panel of Fig.~\ref{figcases3}). This scaling law is also checked by the case~\eqref {caseV1} as well as other potentials.
%


\section{Dynamical Epochs inside Event Horizon}\label{Vdynamics}
Looking inside the horizon of the holographic s-wave superconductor, the authors of Ref.~\cite{Hartnoll:2020fhc} recently studied the dynamical epochs inside the horizon after the spontaneous scalarization appears in a planar asymptotically AdS black hole.  When the temperature is slightly below the critical temperature $T_c$, they found the collapse of the Einstein-Rosen bridge and Josephson oscillations of the condensate. In addition, they also found that geometry near the singularity shows a Kasner cosmology and sometimes with transitions that change the Kasner exponents. Their model is a special case of our general theory by setting $k=0, V(|\Psi|^2)=m^2|\Psi|^2-6$ and $d=2$, \emph{i.e.} a free charged scalar.  Then a question arises naturally: how interactions of the scalar field could alter the interior dynamics as compared to the case of Ref.~\cite{Hartnoll:2020fhc}?

Notice that whether or not the scalarization is spontaneous is not important in the study of dynamical epochs inside the event horizon. Instead of temperature~\cite{Hartnoll:2020fhc}, one can directly use the value of scalar at the event horizon $\psi_H=\psi(z_H)$ to measure the strength of scalar field. Then, we study how the dynamical epochs inside the horizon depend on $\psi_H$. It was noticed that near $T_c$ both the collapse of the Einstein-Rosen bridge and Josephson oscillations happen when scalar field is small. The crucial point in Ref.~\cite{Hartnoll:2020fhc} is that the ``mass term'' in equation of scalar field and the charged coupling term in the Maxwell equation are negligible. As have mentioned by Ref.~\cite{Hartnoll:2020fhc}, it is not simply $\psi$ being small that allows terms to be dropped and the nonlinear dynamics effects always play important roles inside black hole no matter how small the condensation is. Thus, naturally one may wonder if the nonlinear potential $V$ may change these phenomena. This question is not easy to answer analytically due to the complicated forms of Eqs.~\eqref{eqschi0}-\eqref{eqspsi0}. Following the spirit of Ref.~\cite{Hartnoll:2020fhc}, we do that by a ``posteriori assumption'', \emph{i.e.} we first assume that the potential term could be dropped, then find what are the self-consistent conditions and check the conclusions numerically finally. 

\begin{figure}
\centering
  \includegraphics[width=.48\textwidth]{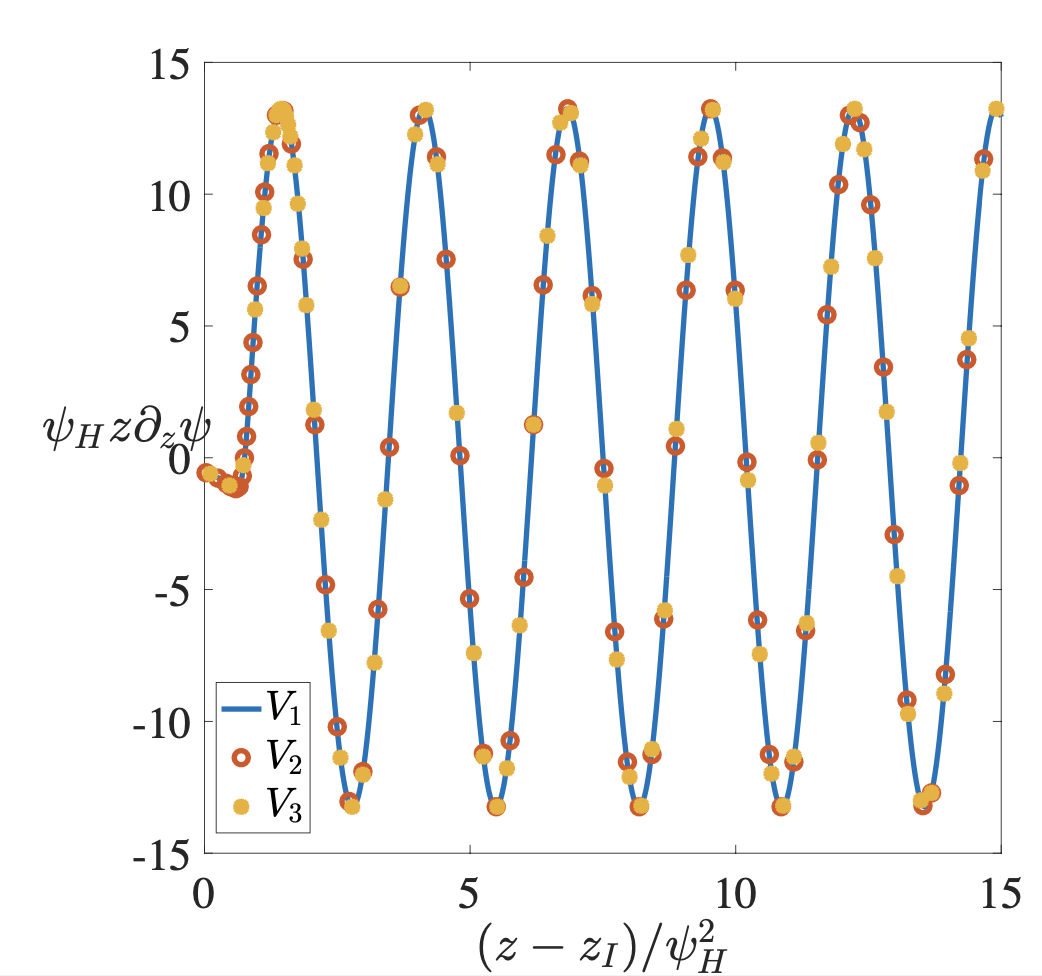}
    \includegraphics[width=.475\textwidth]{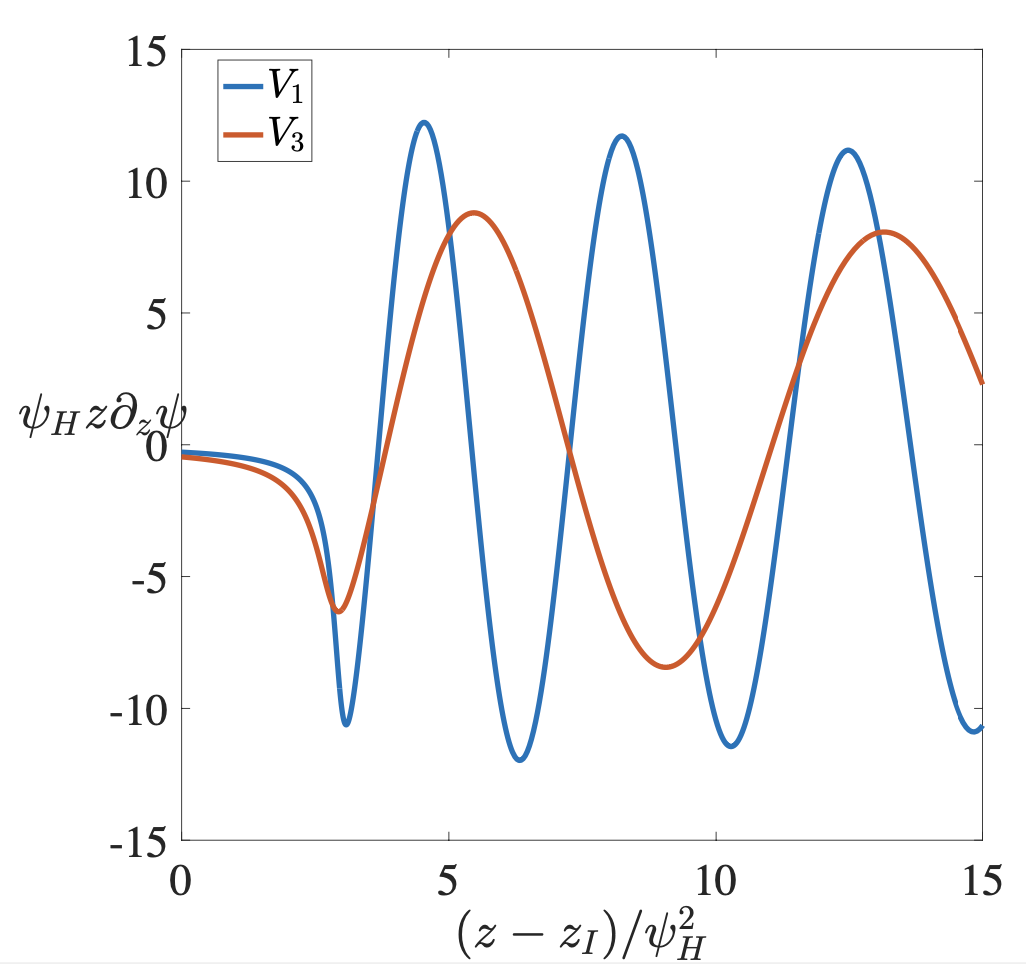}
  \caption{Behaviors of $z\psi'$ around the $z=z_I$ in three different potentials $V_1(x)=x-6, V_2(x)=x+2x^3-6$ and $V_3(x)=-99x+\sinh(100x)-6$. Left panel: $\psi_H=10^{-3}$. Right panel: $\psi_H=0.25$.}
  \label{oscil1}
\end{figure}

Based on the results of Ref.~\cite{Hartnoll:2020fhc}, when we drop above two terms, the scalar field around the ``would-be'' inner horizon $z=z_I$ of RN AdS has the following behavior:
\begin{equation}\label{innerpsi1}
  \psi(z)=\psi_H\cos\left\{\frac{c_0}{\psi_H}\ln[-g_{tt}(z)]+c_1\right\}\,,
\end{equation}
with $c_0$ and $c_1$ two constants. When the scalar field at the event horizon is small, which corresponds to the case that the spontaneous condensation just happens in the model of Ref.~\cite{Hartnoll:2020fhc}, $\psi'$ tends to be divergent and thus leads to strong nonlinear dynamics singularity near the ``would-be'' inner horizon of RN AdS. However, the scalar field itself is bounded and will approach to zero, for which the non-linear terms of $V$ is not important. Thus, we anticipate that our ``posteriori assumption'' is self-consistent for all smooth potentials. This suggests that the collapse of the Einstein-Rosen bridge and Josephson oscillations of the condensate would not depend on the potential $V(|\Psi|^2)$ as long as $\psi_H$ is sufficiently small. To check this conclusion explicitly, we take the following parameters and initial values for numerically solving Eqs.~\eqref{eqschi0}-\eqref{eqspsi0}.
\begin{equation}\label{soluq2}
z_H=Z=q=\tilde{Q}(z_H)=1,~~\chi(z_H)=h(z_H)=A_t(z_H)=k=0,d=2\,.
\end{equation}
There is a Cauchy horizon at $z_I\approx4.9675$ when $\psi(z_H)=0$ (\emph{i.e.} RN AdS case). In Fig.~\ref{oscil1}, we show the value of $z\psi'$ for different potentials.
It is clear that the behavior of $z\psi'$ around the $z=z_I$ are independent of the potential $V$ when $\psi_{H}$ is small (left panel). On the other hand, for large value of $\psi_H$ (right panel), the interactions of the scalar field become important and the collapse of the Einstein-Rosen bridge and Josephson oscillations derivate from the free scalar case. We also found that both the collapse of the Einstein-Rosen bridge and subsequent Josephson oscillations becomes less dramatic as $\psi_H$ becomes large and tend to disappear for sufficiently large vale of $\psi_H$.

We have shown in our main text that the spacetime ends at a spacelike Kasner singularity when the condition~\eqref{constV1} is satisfied. In this case, the potential term can be neglected around the spacetime singularity and so all the analyses of Ref.~\cite{Hartnoll:2020fhc}, including the Kasner inversions, seem to be valid. Unfortunately, due to the complicated forms of Eqs.~\eqref{eqschi0}-\eqref{eqspsi0}, we are not able to understand the dynamical epochs inside event horizon thoroughly so far. Nevertheless, we have shown that, once the condition~\eqref{constV1} is broken, the asymptotic behavior around the spacetime singularity will be different from the Kasner behavior and the Kasner inversion is also broken.

\providecommand{\href}[2]{#2}\begingroup\raggedright\endgroup

\end{document}